\documentclass[aps,superscriptaddress,prx,twocolumn,showpacs,letterpaper,tighten,float,nofootinbib]{revtex4-1}
\usepackage{graphicx}
\usepackage{subfigure}
\usepackage{latexsym}
\usepackage{amsmath,amssymb}
\usepackage{bm} 
\usepackage{color}
\usepackage{stackrel}
\usepackage{epsfig}

\definecolor{myblue}{rgb}{.93, .93, 1}

\newcommand{\bsub}{\begin{subequations}}
	\newcommand{\esub}{\end{subequations}}

\begin{document}

\title{Gapless insulating edges of dirty interacting topological insulators}
\author{Yang-Zhi~Chou}\email{YangZhi.Chou@colorado.edu} \author{Rahul M. Nandkishore} \author{Leo~Radzihovsky}

\affiliation{Department of Physics and Center for Theory of Quantum
  Matter, University of Colorado Boulder, Boulder, Colorado 80309,
  USA} \date{\today}
	
\begin{abstract}
  We demonstrate that a combination of disorder and interactions in a
  two-dimensional bulk topological insulator can generically drive its
  helical edge {\em insulating}. We establish this within the
  framework of helical Luttinger liquid theory and exact Emery-Luther
  mapping. The gapless glassy edge state
  spontaneously breaks time-reversal symmetry in a `spin glass'
  fashion, and may be viewed as a {\it localized} state of solitons which carry half integer charge.  Such a qualitatively
  distinct edge state provides a simple explanation for heretofore
  puzzling experimental observations. This phase exhibits a striking non-monotonicity, with the edge growing less localized in both the weak and strong disorder limits. 
\end{abstract}

\maketitle
	
\section{Introduction}
	
Symmetry protected topological (SPT) phases, of which topological
insulators \cite{Kane2005_1} are the archetypal examples, are ground
states of quantum matter that are `gapped' insulators in the bulk, but
have symmetry-enforced exotic surface properties
\cite{Hasan2010_RMP,SenthilARCMP}.  The conventional wisdom holds that
at the boundary of a topological insulator there exist metallic
surface states, protected by time reversal (TR) symmetry. For
two-dimensional topological insulators
\cite{Kane2005_1,Kane2005_2,Bernevig2006,Hasan2010_RMP},
the metallic edge is expected to support perfect ballistic conduction,
and has been proposed to realize Majorana and $\mathrm{Z}_4$
parafermion zero modes when placed in proximity to a superconductor
\cite{Fu2009,Zhang2014,Orth2015,Alicea2016Review}, opening the door to
entirely new quantum technologies. A more sophisticated understanding
of SPT phases also allows for the possibility of {\it gapped} edges,
as long as the gapped edge either exhibits topological order or breaks
the protecting symmetry \cite{SenthilARCMP}.  Are the above
possibilities of a gapless ballistic or gapped edge exhaustive?

\begin{figure}[t!]
\includegraphics[width=0.3\textwidth]{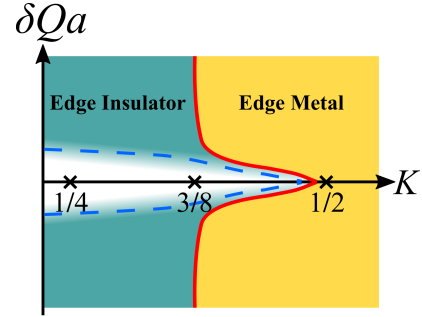}
\caption{The phase diagram of a disordered, interacting edge state of
  a 2D topological insulator. The horizontal axis is the Luttinger
  parameter, $K$, which encodes the interaction. $K=1$ is the
  non-interacting limit and $K=1/4$ is the strongly repulsive, exactly
  solvable Luther-Emery point. The vertical axis is the measure of the
  incommensuration $\delta Q=(4k_F-Q)$ between electron and ion
  densities, with $Q$ the reciprocal lattice vector.  The red solid
  curve denotes the phase boundary between the gapless insulating edge
  (green region) and the helical ballistic edge. The gapless edge
  insulator corresponds to the quantum glass edge state identified in
  this paper, which is a Bose-glass phase \cite{Fisher1989} of helical
  edge bosons. A commensurate gapped insulating edge state (white
  shaded region), which is stable for weak bounded disorder, is
  rendered gapless by Lifshitz tails for strong or unbounded
  disorder.  In the latter case, there are only two phases, a glassy
  insulating edge and the helical Luttinger liquid edge, with a crossover
  inside the glassy phase (blue dashed line) which is further
  illustrated in Fig.~\ref{Fig:LE_DoS}.}
\label{Fig:TBreaking}
\end{figure}
	
In fact, experiments suggest a richer set of possibilities
\cite{Konig2007,Knez2011,Suzuki2013,Du2015,Li2015,Qu2015,Nichele2016,Nguyen2016,Couedo2016,Fei2017,Du2017,Ma2015}.
Specifically, while short and intermediate length samples of
two-dimensional topological insulator indeed exhibit quantized
ballistic transport, longer samples show conductance well below
$e^2/h$ per edge \cite{Konig2007,Du2015,Du2017}.  Furthermore,
insulating transport was reported in InAs/GaSb \cite{Li2015,Nichele2016}, whereas
InAs/GaSb \cite{Du2015} and HgTe/CdTe \cite{Ma2015} show metallic
transport that appears to be robust to time-reversal (TR) symmetry
breaking external magnetic field. This rich experimental phenomenology
motivates a re-examination of the dogma that two dimensional
topological insulators must have either a gapped or a perfect metallic
edge.
	
In this article, we show, using nonperturbative analyses on a minimal model, that two
dimensional topological insulators can exhibit a third possibility: a
{\it gapless, insulating} edge. This possibility - which runs counter
to prevailing wisdom on SPT phases in general and topological
insulators in particular - becomes available through an interplay
between disorder and interactions. Given that theoretical analyses
seldom simultaneously treat both disorder and interactions \cite{Kane1992PRL,Kane1992PRB,Kane1997}, it is
unsurprising that this possibility has not been emphasized. Nevertheless,
{\it experimental} systems are invariably both disordered and
interacting, and thus for real materials, gapless insulating edges are
a generic possibility.
	
We focus on the boundary of a two dimensional TR-invariant topological
insulator, which we model by helical Luttinger liquid theory
\cite{Kane2005_2,Wu2006,Xu2006}. This provides a natural (and
nonperturbative) way to incorporate the effect of short-range
interactions. We then study the effect on the interacting edge states
of TR-invariant disorder. This problem is very different from the
conventional disordered spinless Luttinger liquid, where Anderson
localization dominates the low temperature physics, in that the famous
`topological protection' of surface states \cite{Hasan2010_RMP}
forbids conventional single particle backscattering. As a result, a
minimal model of the 2D topological insulator edge state only contains
perturbations from the forward scattering disorder \cite{Xie2016} and
the umklapp interaction.  Forward scattering disorder alone has no
effect on transport (it generates nontrivial Luttinger parameter
$K$). Meanwhile, the umklapp interaction can spontaneously break TR
symmetry and open a gap at special (commensurate) fillings, but is
irrelevant in the renormalization group (RG) sense at generic
(incommensurate) filling. Thus, under generic conditions,
neither forward scattering nor umklapp interactions {\it alone} should
affect transport.
	
However, as we will show below, forward scattering and umklapp
interactions {\it together} give rise to a gapless, insulating
edge. In essence, umklapp interaction produces the backscattering,
which disorder alone is `topologically prohibited' from doing, while
disorder locally compensates for the momentum mismatch
(incommensuration). As a result, the {\it combination} of disorder and
interactions can accomplish what neither can alone, giving rise to an
entirely new state on the edge, which is gapless and insulating. This
phase {\it locally} breaks TR symmetry, but in a `spin glass'
\cite{Edwards1975} fashion, where the `sign' of the TR symmetry
breaking order parameter is spatially random. It preserves {\it
  statistical} TR symmetry, (i.e. after spatial or disorder
averaging), but is {\it localized} \cite{pwa1958} and therefore
insulating. Being localized, this state is stable to non-zero energy densities, in a manifestation of localization protected order \cite{LPQO}, with the added subtlety that the order is itself required to enable localization \cite{LRMBL}.
	
The possibility of glass-like TR breaking in the ground state was
anticipated already in Ref.~\onlinecite{Wu2006,Xu2006}. However, they focused on non-generic commensurate filling, at which interactions gap out the edge (as explicitly predicted in Ref.~\onlinecite{Wu2006}), thereby precluding the {\it gapless} insulating edge predicted here. 
While the possibility of a gapless insulating edge follows naturally from the observations in Ref.~\onlinecite{Wu2006,Xu2006}, as far as we are aware it has not been explored in the literature. 

Here we explore the possibility of a gapless insulating edge and its phenomenology via Luttinger liquid and exact Luther-Emery analyses. Our basic strategy is to (i) use bosonization to demonstrate an instability of the Luttinger liquid for $K < 3/8$ (generalizing the arguments of Refs. \onlinecite{Wu2006,Xu2006} to incommensurate filling), and (ii) to then infer the properties of the system for $K < 3/8$ by combining the bosonized analysis with an exact solution of the problem (using refermionization) at the Luther-Emery point $K=1/4$. This strategy will be valid as long as there are not any additional phase transitions for $1/4<K<3/8$. ) In this manner, we arrive at the following key results: (a) the edge is a ``non-Fermi" glass \cite{Parameswaran_NF_glass} best thought of as a localized state of edge solitons with half-integer charge see Sec.~\ref{Sec:QuSG}, (b) the localization length exhibits a striking non-monotonic dependence on
the strength of disorder, predicting weakening of localization at both weak and
strong disorder see Sec.~\ref{Sec:QuSG}, and (c) a distinctive phenomenology (for magnetic field response and unexpected resistance), providing a natural interpretation of a number of current experiments on edge transport in topological insulators \cite{Du2015,Li2015,Du2017,Li2017}. An extensive discussion of the implications of our results for experiments is provided in Sec.~\ref{Sec:discussion}. Readers uninterested in the technical details may skip directly to this section.

\section{Model}
	
At the edge of a two-dimensional topological insulator, there arise
counter-propagating states of right $(R)$ and left ($L$) moving
fermions, that are helicity eigenstates. At low energies, the `kinetic
energy' part of the Hamiltonian takes the form
\begin{align}\label{Eq:H_0}
  \hat{H}_0=-i v_F\int dx
  \left[R^{\dagger}(x)\partial_xR(x)-L^{\dagger}(x)\partial_xL(x)\right],
\end{align}
where $v_F$ is the Fermi velocity. 
This Hamiltonian possesses an anti-unitary time-reversal symmetry
under which $R(x)\rightarrow L(x)$, $L(x)\rightarrow -R(x)$, and
$i\rightarrow -i$, encoding the underlying spin-1/2 structure. This
symmetry rules out conventional backscattering (e.g.,
$R^{\dagger}L+L^{\dagger}R$ in the spinless Luttinger liquid)
\footnote{Technically, impurity backscattering terms with extra
  derivatives are allowed by TR operation. Such terms do arise in the
  presence of Rahsba spin orbit coupling
  \cite{Strom2010,Budich2012,Schmidt2012}, but they do not modify the
  ballistic transport by themselves \cite{Kane2005_1,Xie2016}.}.
Disorder thus gives rise to purely `forward scattering', which
adds to the Hamiltonian a term
\begin{align}\label{Eq:H_V}
  \hat{H}_V=\int dx\,
  V(x)\left[R^{\dagger}(x)R(x)+L^{\dagger}(x)L(x)\right].
\end{align}
For analytical convenience, we take the potential $V(x)$ to be a zero
mean Gaussian random field, fully characterized by
$\overline{V(x)V(y)}=\Delta\delta(x-y)$, where
$\overline{\mathcal{O}}$ denotes a disorder average of
$\mathcal{O}$. Additionally, short-range interactions give rise to
two-particle umklapp backscattering (consistent with TR
symmetry),
\begin{align}
  H_U=U\int dx \left[ e^{-i\delta Q x}
    L^{\dagger}(x+\alpha)L^{\dagger}(x)R(x)R(x+\alpha)
    +\text{H.c.}  \right],
\end{align}
where a point splitting with the ultraviolet length $\alpha$ is performed.
Here $U>0$ is the strength of the two-particle backscattering
interaction and
\begin{align}
  \delta Q=Q-4k_F
\end{align}
measures the mismatch (lack of commensuration) between electron and
ion density, $Q=2\pi/a$ with $a$ the lattice constant of the two dimensional bulk
topological insulator.

Employing standard bosonization
\cite{Shankar1995,Giamarchi_Book,Shankar_Book} to treat the
interaction nonperturbatively, the system is characterized by an
imaginary-time action
$\mathcal{S}=\mathcal{S}_0+\mathcal{S}_V+\mathcal{S}_U$ for a
phonon-like field $\theta$, where
\begin{subequations}\label{Eq:S}
 \begin{align}
   \label{Eq:S_0}\mathcal{S}_0=&\int d\tau dx
\,\frac{1}{2\pi vK}\left[\left(\partial_{\tau}\theta\right)^2
+v^2\left(\partial_x\theta\right)^2\right],\\
\label{Eq:S_V}\mathcal{S}_V=&\int d\tau dx \,V(x)
\frac{1}{\pi}\partial_x\theta,\\
\label{Eq:S_U}\mathcal{S}_U=&\tilde{U}\int d\tau dx
\,\cos\left[4\theta-\delta Qx\right],
\end{align}
\end{subequations}
with $v$ the velocity of the boson, $K$ the Luttinger parameter,
$\tilde{U}=U/(2\pi^2\alpha^2)$, and $\alpha$ an ultraviolet length
scale \footnote{A generic model would also include a (TR symmetry
  allowed) random two particle backscattering, leading to a short
  range correlated random contribution to the coefficient $\tilde U$
  in Eq.~(\ref{Eq:S_U}). As long as $\tilde U(x)$ has non-zero mean
  (arising from interactions) the behavior remains qualitatively the
  same.}.  For a system with repulsive interactions $K < 1$ ($K=1$
being the non-interacting point).  The bosonized form of the
long-lengthscale part of electron density is given by
$n=\partial_x\theta/\pi$.  We take the disorder and interactions to be
sufficiently weak, that they do not close the {\em bulk} gap, i.e.,
the bulk topological insulator phase is stable.
	
\section{Quantum Glass State}\label{Sec:QuSG}
	
\begin{figure}[t!]
\includegraphics[width=0.4\textwidth]{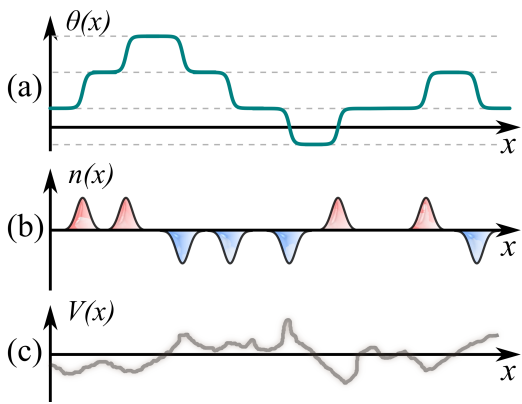}
\caption{Domain wall proliferation in the presence of disorder. (a)
  The configuration of $\theta(x)$. (b) The associated density
  profile. (c) The disorder potential.  The value of $\theta$ is
  restricted to $\theta=(2N+1)\pi/4$ (integer $N$) for
  minimizing the $\mathcal{S}_U$ [given by Eq.~(\ref{Eq:S_U})] in the
  clean case.  In the presence of disorder, frozen-in domain walls
  appear, corresponding to solitons (red shaded density bump) and
  anti-soliton (blue shaded density depletion), and lead to spatial
  wandering of $\theta$ between equivalent minima.  These excitations
  carry half of an electron charge \cite{Qi2008}.  The solitons and the
  anti-solitons are ``pinned'' by disorder potential $V(x)$ [sketched
  in (c)], leading to the quantum 'spin-glass' edge state, which may be viewed as a 
localized state of these half-charge solitons.}
\label{Fig:DW}
\end{figure}

\begin{figure}[t!]
	\includegraphics[width=0.4\textwidth]{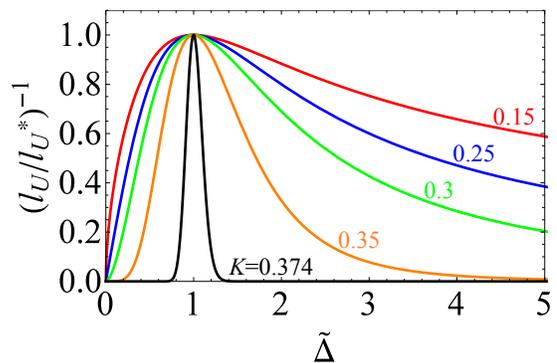}
\caption{The inverse localization length as a function of disorder. The localization length $l_U$ shows non-monotonic behavior as a function of the dimensionless parameter, $\tilde{\Delta}=8K^2\Delta/(v^2\delta Q)$, which measures disorder strength.
Here $l_U$ is extracted via RG analysis, $l_U\sim\alpha\Delta_U^{-1/(3-8K)}$, and $l_U^*$ is the smallest value of the localization length with fixed values of $\delta Q$ and $\tilde{U}$. Different values of the Luttinger parameters are plotted in black ($K=0.374$), orange ($K=0.35$), green ($K=0.3$), blue ($K=0.25$), and red ($K=0.15$) curves. 
Regardless of the interaction, the localization length diverges both in the zero disorder limit and in the infinite disorder limit. 
}
\label{Fig:loc_L}
\end{figure}

In the clean system [$V(x)=0$] with commensuration ($\delta Q=0$), the
two-particle backscattering in Eq.~(\ref{Eq:S_U}) is relevant for
$K<1/2$ \cite{Giamarchi_Book}. It spontaneously breaks the TR symmetry
at zero temperature \cite{Wu2006}, and opens up a gap at the edge.
However, (a) commensuration is unlikely in typical samples, requiring
fine-tuning, (b) spontaneous ordering of this sort will not survive to
non-zero temperatures \cite{Peierls}, and (c) general arguments
\cite{Imry1975} establish that such a long-range ordered state is
unstable to arbitrarily weak disorder in one dimension.
	
\begin{figure}[t!]
  \includegraphics[width=0.35\textwidth]{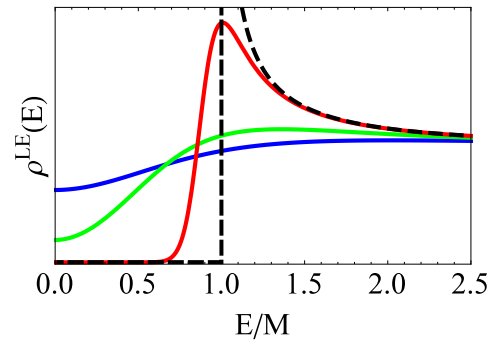}
  \caption{The disorder-averaged density of states of Luther-Emery
    fermions with Hamiltonian Eq.~(\ref{Eq:LE_H}) based on the
    analytic solution in Ref.~\onlinecite{Bocquet1999}, for a range of variances,
    $\tilde{\Delta}/(Mv)=$0.1, 0.5, 1 (red, green, blue respectively)
    of Gaussian disorder potential.  The black dashed line is the
    disorder-free density of states. For unbounded disorder (e.g.,
    Gaussian) distributions, the density of states is nonzero for
    arbitrarily small $\Delta$, due to exponentially small subgap
    Lifshitz tails, a relic of a Mott insulator in the clean edge.}
  \label{Fig:LE_DoS}
\end{figure}

In the absence of disorder, but at incommensurate electron density
($\delta Q\neq 0$), weak two-particle backscattering, $\mathcal{S}_U$
is formally irrelevant due to kinematic constraints from momentum
conservation.  The symmetry and the gapless edge will be restored by
sufficiently large incommensuration $|\delta Q| > \delta Q_c$, via a
commensurate-incommensurate phase transition \cite{PokrovskyTalapov}.
In the presence of disorder, however, the kinematic constraint is
relaxed, and the backscattering can be enhanced.  To treat the
combination of interaction and disorder rigorously, we first perform a
change of variable, $\theta(x)\rightarrow
\theta(x)-\frac{K}{v}\int_{-\infty}^{x}ds V(s)$, to
eliminate the $\mathcal{S}_V$ term. Then $\mathcal{S}_U$ becomes
\begin{subequations}\label{eq:U'}
\begin{align}
\mathcal{S}_U=&\frac{\tilde{U}}{2}\int d\tau dx\left\{
\eta(x)\,e^{i4\theta(\tau,x)}+\eta^*(x)\,e^{-i4\theta(\tau,x)}
\right\}\\
=&\tilde{U}\int d\tau dx\,\cos\left[4\theta(x)+\chi(x)\right],\label{eq:U'2}
\end{align}
\end{subequations}
where $\eta(x)=e^{i\chi(x)}$,
$\chi(x)=-\frac{4K}{v}\int_{-\infty}^{x}ds V(s)-\delta Qx$. 
In the thermodynamic limit and with a non-zero $\Delta$,
the random field $\eta(x)$ is well characterized by its second cumulant
$\overline{\eta^*(x')\eta(x)}=e^{-\frac{8K^2}{v^2}\Delta|x-x'|}e^{-i\delta
 Q(x-x')}$, $\overline{\eta(x')\eta(x)}=0$, and a vanishing mean $\overline{\eta(x)}=0$.
Higher cumulants of $\eta(x)$ are nonzero, but lead to
higher replica operators that are less relevant and can be neglected
relative to the second cumulant that we kept. (A discussion on the behavior of finite sized systems can be found in Appendix~\ref{App:RG}.)
Here $\mathcal{S}_U$ given by Eq.~(\ref{eq:U'}) is an umklapp two particle backscattering with a
position-dependent random phase $\chi(x)$ executing a random walk
\footnote{The random two particle backscattering terms were first
  obtained from symmetry analysis in
  Ref.~\onlinecite{Wu2006,Xu2006}. Note however that the
  backscattering term in Eq.~(\ref{eq:U'2}) contains a uniform
  amplitude, which is different from those in
  Ref.~\onlinecite{Wu2006,Xu2006}.}.  This effective random-field XY
model maps to the Bose-glass problem analyzed in \cite{Giamarchi1988,
  Fisher1989}, and supports a glassy (i.e. insulating) phase. In the
presence of Gaussian disorder, the gap will be smeared out, although a
strong crossover controlled by Lifshitz tails will survive at weak
disorder, illustrated in Figs.~\ref{Fig:TBreaking}
and~\ref{Fig:LE_DoS}.

To treat the disordered-assisted umklapp action $\mathcal{S}_U$ [given by Eq.~(\ref{eq:U'})] we
employ a replica method \cite{Edwards1975} (equivalent to the Keldysh
formalism), which allows us to average over disorder, generating a
replicated (effectively) translationally-invariant action (see Ref.~\onlinecite{Kainaris2014} and Appendix~\ref{App:RG}),
\begin{align}\label{Eq:S_U_dis}
  \nonumber\mathcal{S}_{U,dis}=&-\Delta_U\sum_{a,b}\\
  &\times\int d\tau d\tau' dx
  \cos\left[4\left(\theta_a(\tau,x)-\theta_b(\tau',x)\right)\right].
\end{align}
Here,
\begin{equation}
\Delta_U=\tilde{U}^2\frac{K^2\Delta/v^2}{16(K^2\Delta/v^2)^2+\delta \label{eq: nonmonotonic}
  Q^2/4},
\end{equation}
and $a,b$ are replica indices, with the standard zero-replica limit to
be taken at the end of the computation.  Note that the argument of the
cosine in Eq.~(\ref{Eq:S_U_dis}) is insensitive to
incommensuration. Intuitively, disorder takes care of the
commensuration by `supplying the missing momentum' to make the
interaction locally commensurate. We note in passing that this `positive feedback' has been previously
discussed in the context of finite temperature transport in the
perturbative regime \cite{Fiete2006,Kainaris2014,Chou2015}.

Although the physical origin of the disorder-assisted interaction is
quite different,
$\mathcal{S}_{U,dis}$, is formally identical to the random
single-particle backscattering in the Giamarchi-Schulz model, with
rescaling $\theta \rightarrow 2 \theta$, corresponding to
$K\rightarrow 4K$ \cite{Kainaris2014}.  It follows from a standard
analysis that this operator becomes relevant for $K < 3/8$
\cite{Giamarchi1988,Wu2006,Xu2006}, driving an instability to a Bose-glass phase,
corresponding to a gapless localized edge, as we discuss below. An
estimate for the localization length in the glass phase may be
obtained from the RG analysis \cite{Giamarchi1988,Giamarchi_Book} that
predicts a length scale $l_U\sim\alpha \Delta_U^{-1/(3-8K)}\propto
U^{-2/(3-8K)}$ at which the disorder assisted umklapp becomes
strong. We further note that the localized nature of the edge can
stabilize order to non-zero energy densities \cite{LPQO}.

To discuss the nature of the glassy edge state we note that
forward-scattering disorder forces $\theta(x)$ to jump between
degenerate minima of Eq.~(\ref{Eq:S_U}) whenever $V(x)$
locally exceeds the critical $\delta Q_c$ (corresponding to the
soliton gap), thereby producing a random distribution of domain walls,
as illustrated in Fig.~\ref{Fig:DW}
(for a detailed discussion see Appendix~\ref{App:DW_dis}). Note that
domain walls connect two states related by TR operation, and are characterized
by a $\pi/2$ change in $\theta$. Since the charge density is
$\frac{e}{\pi} \partial_x \theta$, it follows that the domain walls are
fractionalized charge $e/2$ excitations \cite{Qi2008,Ziani2015}. The glassy edge
state is best thought of as a `localized state of half-charge soliton' rather
than of bare fermions, and as such may be viewed as an example of a
`non-Fermi glass' \cite{Parameswaran_NF_glass}.  Note that such a
localized state of domain walls spontaneously breaks TR symmetry in a
spatially random fashion.

We emphasize that effective strength of the disorder assisted umklapp
backscattering $\Delta_U(\Delta)$ in Eq.~(\ref{eq: nonmonotonic}) is a
non-monotonic function of $\Delta$. 
(Also see Fig.~\ref{Fig:loc_L} for inverse localization length.) 
Clearly, the effects of disorder
increase with $\Delta$ at small disorder, with the edge being
ballistic in the zero disorder limit. Meanwhile, it follows from
inspection of Eq.~(\ref{Eq:S_V}) that $V(x)$ (viewed as a smoothly
varying function) locally increases incommensuration (adding a random
contribution to $\delta Q$ in Eq.~(\ref{Eq:S_U}), thereby locally
proliferating domain walls). For typical $V(x) > v\delta Q$ this has
the effect of suppressing umklapp backscattering. Since $\Delta$
represents the typical strength of $V(x)$, it follows that
localization should paradoxically get {\it weaker} as $\Delta$ is
increased in the strong disorder regime.

A more rigorous treatment of the strong-coupling phase may be obtained
by re-fermionizing the model given by Eq.~(\ref{Eq:S}) at the
Luther-Emery point ($K=1/4$) with a standard transformation of $\theta$ (see Ref.~\onlinecite{Haldane1981}, the section 4.2 of Ref.~\onlinecite{Giamarchi1988}, and Appendix~\ref{App:LE_mapping}).
The corresponding Luther-Emery Hamiltonian is given by
\begin{align}
\nonumber\hat{H}_{LE}=&-iv\int dx\left[\Psi^{\dagger}_R\partial_x\Psi_R-\Psi^{\dagger}_L\partial_x\Psi_L\right]\\
\nonumber&+M\int dx \left[\Psi^{\dagger}_R\Psi_L+\Psi^{\dagger}_L\Psi_R\right]\\
\label{Eq:LE_H}&+\frac{1}{2}\int dx [V(x)+ v \delta Q] \left[\Psi^{\dagger}_R\Psi_R+\Psi^{\dagger}_L\Psi_L\right],
\end{align}
where $M=U/(2\pi\alpha)$ is the mass of Luther-Emery fermion.
Note that in this representation, the problem has simplified to a non-interacting theory with disorder.

The Luther-Emery fermion obeys a massive Dirac equation with a
spatially inhomogeneous scalar potential $V(x)/2$.  The corresponding
variance of the scalar potential $V(x)/2$ is
$\tilde{\Delta}=\Delta/4$.  The problem of a non-interacting massive
fermion with scalar potential disorder is exactly solvable
\cite{Bocquet1999}, and exhibits localization for arbitrarily weak
disorder at all energies (i.e. for arbitrary incommensuration $\delta
Q$). For unbounded disorder, there is no gap in the density of
states. Nevertheless, in the weak disorder limit
$\tilde{\Delta}/(Mv)\ll 1$, the subgap density of states is
exponentially small, whereas for strong disorder
$\tilde{\Delta}/(Mv)\gg 1$, the gap is completely smeared out, leading
to a crossover driven by disorder strength which is illustrated in
Fig.~\ref{Fig:LE_DoS}. For weak bounded disorder, a distinct gapped
phase survives.
	
The physical density and current operators can be written in terms of
the Luther-Emery fermion fields:
$n=\frac{e}{2}\left(\Psi_R^{\dagger}\Psi_R+\Psi_L^{\dagger}\Psi_L\right)$
and
$j=\frac{ev}{2}\left(\Psi_R^{\dagger}\Psi_R-\Psi_L^{\dagger}\Psi_L\right)$,
where $e<0$ is the charge of an electron. Since these fermions (which
carry half integer charge) are localized, it follows that the physical
conductivity vanishes. Meanwhile, the absence of a hard gap for the
Luther-Emery fermions implies that the physical compressibility is
always nonzero. Thus the edge is a localized compressible state of
Luther-Emery fermions, which in turn correspond to domain walls of
$\mathcal{S}_0+\mathcal{S}_U$ [given by Eq.~(\ref{Eq:S})] in the
language of bosonization. We thus conclude, that at $K=1/4$,
consistent with the bosonization arguments there arises a localized
state of half-charge solitons (whose densities are govern by domain walls), for arbitrary filling $\delta Q$ and even at
weak disorder, which generically manifests as a gapless insulating
edge.
	
Note that $V(x)$ in Eq.~(\ref{Eq:LE_H}) plays a dual role: it both
introduces disorder and acts as a local chemical potential, increasing
the density of effective fermions. Away from the Luther-Emery point $K
= 1/4$, the effective fermions are interacting (see Appendix~\ref{App:LE_mapping} for a discussion), and the Anderson
insulator grows less stable to interactions with increasing
density. Again, the gapless insulating state is expected to be stable for $K<3/8$. 
It then follows that increasing disorder beyond an optimal
value set by $\delta Q$ strikingly weakens localization on the
edge. This provides a complementary perspective on the aforementioned
non-monotonicity [Eq.~(\ref{eq: nonmonotonic}) and Fig.~\ref{Fig:loc_L}] of the effects of
disorder.

\section{Experimental signatures of quantum glass edge}

Having established the insulating glass nature of the disordered
interacting edge of the topological insulator, many experimental
predictions then follow from general understanding of randomly-pinned
and driven elastic media, (e.g., charge-density waves
\cite{Fukuyama1978,Narayan1992,Giamarchi1996,Balents1998,Gorkov_cdw_book}),
and quantum transport in the localized state \cite{Imry2002}.

Owing to the glassy insulating nature of the edge, at zero
temperature, a finite length $L$ sample will be characterized by a dc
conductance $G\sim (e^2/h)e^{-L/l_{U}}$ per edge, vanishing in
thermodynamic limit. We expect the corresponding ac conductivity to be
described by a Mott form $\sigma(\omega)\sim \omega^2$ for low
frequency, and non-universal power law dependence $\sigma(\omega)\sim
\omega^{-4+8K}$ for the high frequency tail
\cite{Giamarchi1996}. 

Quite generically we therefore expect that at zero temperature, above
a threshold electric field, the glassy edge will exhibit a collective
depinning transition from an insulating to a conducting state, studied
extensively in the context of charge-density waves
\cite{Fukuyama1978,Narayan1992,Giamarchi1996,Balents1998,Gorkov_cdw_book}. We
thus predict a nonlinear I-V characteristic for the low-temperature
edge transport. Concomitantly, we expect above-threshold transport to
exhibit narrow-band noise and mode-locking phenomena at characteristic
frequency $\omega\sim e V/\hbar$, set by voltage, $V$.
	
In addition to transport measurements, there are a couple of
experimental protocols that reveal the properties of quantum glass
states.  Compressibility measurement can confirm the
gaplessness. Scanning tunneling microscopy, which measures the local
density of states, can tell a gapless state and provide spatial
distribution of electrons in a fixed energy.  The half charge nature
can in principle be probed through Coulomb blockade  \cite{Qi2008}.
Charge density autocorrelation functions on the edge should find
glassy dynamics.  Non-local charge response \cite{Khemani2015} on the
edge also gives distinct signatures of the localized state.

\section{Discussion and Conclusion}\label{Sec:discussion}
	
We have shown that an interacting edge of a topological insulator with
symmetry-preserving disorder can exhibit a quantum glass phase that
spontaneously breaks time-reversal symmetry.  This phase is a gapless
compressible insulator that corresponds to a Bose-glass of the
phonon-like field $\theta$ \cite{Fisher1989}, or equivalently a
localized state of half-charge solitons. This constitutes a
qualitatively new possibility for the edge of a two-dimensional
topological insulator (and more generally, SPT phases), distinct from
a metallic or a gapped edge. Insofar as this state is localized, it
can survive to non-zero energy density.
The corresponding phase diagram is sketched in
Fig.~\ref{Fig:TBreaking}.
	
\begin{figure}[t!]
\includegraphics[width=0.45\textwidth]{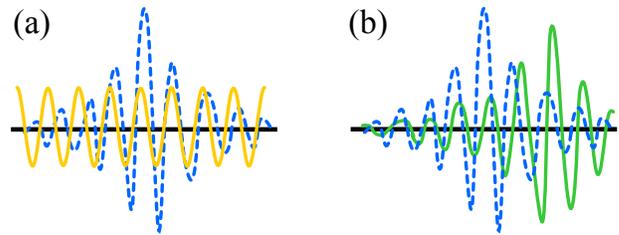}
\caption{Schematic illustration of the modification of the
  wavefunction in the presence of external magnetic field. (a) The
  sketched wavefunctions for $K>3/8$. The yellow curve represents a
  delocalized wavefunction which corresponds to ballistic
  transport. In the presence of magnetic field, the wavefunction
  becomes localized (blue dash curve).  (b) The sketched wavefunction
  for $K<3/8$. In the absence of magnetic field, the ground state is a
  quantum glass (green curve), which is a gapless TR breaking
  insulator. The state remains localized upon turning on the magnetic
  field (blue dash curve).  }
\label{Fig:Mag_wavefcn}
\end{figure}

This new phase provides an alternative perspective on existing
experiments \cite{Knez2011,Du2015,Li2015,Du2017,Li2017}, as we will
now discuss. In order to make a meaningful comparison to experiments,
we must first discuss the evolution of the glassy edge state in
applied magnetic field. How is the transport behavior altered if we
explicitly break TR symmetry by applying external magnetic field?
Inside the glassy phase the edge is already localized and TR symmetry
is spontaneously broken. Thus, applying weak external field has little
effect - it simply shifts the localization length, as shown in
Fig.~\ref{Fig:Mag_wavefcn} (b). In contrast, application of magnetic
field to the TR invariant metallic state triggers a phase transition
to an insulator [see Fig.~\ref{Fig:Mag_wavefcn} (a)], which in the
non-interacting limit is simply an Anderson insulator of electrons
\cite{Maciejko2010}.  A detailed analysis of the effects of applied
field within the language of bosonization is provided in Appendix~\ref{App:B_field}. 
We summarize here the key results.

Application of magnetic field $B$ to the metallic TR preserving state
leads to the {\it typical} conductance 
\footnote{The typical value of conductance ($G$) is defined by $\exp\left(\overline{\ln G}\right)$. The typical conductance and averaged conductance might give quantitatively different values owing to strong fluctuating nature of one dimensional localized insulators \cite{Beenakker1997}.}
\begin{align}\label{Eq:G_HLL}
  G= \frac{e^2}{h}e^{-L/l_B},
\end{align}
where $l_B\propto B^{-2/(3-2K)}$, recovering ballistic transport
when $L\ll l_B$, where $L$ is the length of the sample.  Application
of magnetic field to the glassy TR breaking state proposed here
(stable for $K<3/8$ at generic incommensurate electron density) leads
to the conductance
\begin{align}\label{Eq:G_QG}
  G= \frac{e^2}{h}e^{-L/l_{\text{loc}}},
\end{align}
with $l_{\text{loc}}=\min\left(l_B,l_U\right)$. This crosses over to
Eq.~(\ref{Eq:G_HLL}) only for $B > B_*$ (when $l_B < l_U$), where
\begin{equation}
  B_*\sim l_U^{K-3/2}\sim \Delta_U^{\frac{3/2-K}{3-8K}}
\end{equation}
is the crossover magnetic field below which the system is insensitive
to magnetic field $B$.

How do these predictions compare with experiment? In InAs/GaSb
insulating behavior is observed in the absence of external magnetic
field \cite{Li2015,Nichele2016}. The samples are strongly interacting with
estimated Luttinger parameter $K\sim 0.2$ \cite{Li2015} owing to the
small Fermi velocities \cite{Knez2011,Du2015}. In this $K<3/8$ regime
our theory indeed predicts a gapless insulating edge, consistent with
these experimental observations \footnote{The analyses in this work strictly apply to long edge samples. 
We expect qualitatively similar results for short edge samples, although an effective single impurity picture might be more helpful in this regime \cite{Maciejko2009,Vayrynen2016}.}.

In InAs/GaInSb samples \cite{Du2017,Li2017}, conducting behavior is
observed, but with conductance less than $e^2/h$ for long samples in
the absence of applied magnetic field. In the presence of applied
magnetic field, resistivity increases sharply. In these samples the
estimated Luttinger parameters are roughly 0.4-0.5
\cite{Du2017,Li2017}, close to but greater than our critical value
$K=3/8=0.375$. Thus, we do expect conducting behavior in these
samples, with a transition to insulating behavior driven by applied
magnetic field. We conjecture that the less than ballistic transport
observed in these samples (in the absence of applied field) may be due
to rare regions with $K<3/8$ \footnote{Spatially inhomogeneous
  Luttinger parameter can arise in the presence of random Rashba spin
  orbit coupling \cite{Xie2016}.}, which locally break the
time-reversal symmetry spontaneously and provide finite resistivity.
Of course, the TR symmetry is still approximately restored for any
finite temperature \cite{Wu2006} when the system is coupled to an
external heat bath (e.g. phonons).
	
We note that conducting edges were reported in HgTe/CdTe \cite{Ma2015}
even in the presence of magnetic field. This material is believed to
be weakly interacting ($K\sim 0.8$ \cite{Teo2009}).  This observation
does not have a natural explanation within our theory, but might be
understood by material specific issues
\cite{Li2017_Hidden_edge,Skolasinski2017}.

We comment briefly on some of the other explanations that have been
proffered for the experiments. 
One explanation that has been advanced for resistive behavior \cite{Nichele2016} is that the system may simply be in the trivial phase, without a topologically protected edge. Even if the sample is in the topological phase, 
strong interaction can open up a gap and spontaneously
break TR symmetry \cite{Wu2006}, leading to an insulating behavior. However, 
this requires fine-tuning to commensuration and assumes a
disorder-free system. There are also theories that are specific to
particular material realizations
\cite{Pikulin2014,Hu2016,Li2017_Hidden_edge,Skolasinski2017}.  Unlike
all these, our analysis indicates that the TR-breaking edge state is a
robust and generic feature of the strongly interacting 2D topological
insulator edge states.

Another possibility that has been discussed in the literature is
charge puddles near the edge \cite{Vayrnen2013,Vayrnen2014}. These can
behave like a Kondo impurities \cite{Vayrnen2014,
  Maciejko2009,Tanaka2011} and can generate insulator-like
finite-temperature conductivity \cite{Vayrynen2016}. 
An extensive number of the Kondo impurities can also
localize the helical edge state by spontaneously breaking the
time-reversal symmetry \cite{Altshuler2013,Hsu2017}. Our mechanism bears some
family resemblance to these, but differs crucially in that there is no
need to invoke additional gapless states (such as charge puddles) in
the {\it bulk}, nor to invoke effectively magnetic impurities - our
scenario applies even to the pure one dimensional edge of a
topological insulator that is fully gapped in the bulk, with purely
TR-preserving disorder on the edge. We conclude that our theory of a
gapless insulating edge provides a {\it simple} explanation for a
number experimental observations, independent of material and bulk
details, and sensitive only to the edge theory.
As a non-trivial {\it prediction} of our theory, the glassy edge state
that we identify should exhibit a non-monotonic dependence on disorder
strength, being maximally insulating (i.e. having shortest
localization length) at {\it intermediate} disorder strengths
(Fig.~\ref{Fig:loc_L}). Moreover the localized objects are solitons
with half integer charge.

Thus, we have shown that topological
insulators {\it can} have gapless insulating edge states. This
scenario also provides a simple explanation of apparent experimental
anomalies. While our present analysis was restricted to topological
insulators in two dimensions, a generalization to higher dimensional
systems and more general SPT phases should follow {\it mutatis
  mutandis}. We leave this, as well as a re-examination of other
experiments through this lens, as a challenge for future work.

\section*{Acknowledgements}
	
We thank Ravin Bhatt, Leonid Glazman, Charles Kane, Jed Pixley, Michael Pretko, Abhinav Prem, and Hong-Yi Xie for useful discussions and feedback. 
This work is supported in part by
a Simons Investigator award from the Simons Foundation (Y.-Z.C. and L.R.), and in part by the Army Research
Office under Grant Number W911NF-17-1-0482 (Y.-Z.C. and R.N.). The
views and conclusions contained in this document are those of the
authors and should not be interpreted as representing the official
policies, either expressed or implied, of the Army Research Office or
the U.S. Government. The U.S. Government is authorized to reproduce
and distribute reprints for Government purposes notwithstanding any
copyright notation herein.

	\appendix
	
	\section{Bosonization Convention}
	
	We adopt the standard field theoretic bosonization method
	\cite{Shankar1995,Shankar_Book}. The fermionic fields can be described
	by chiral bosons via
	\begin{align}
	R(x)=\frac{1}{\sqrt{2\pi\alpha}}e^{i\left[\phi(x)+\theta(x)\right]},\,\,L(x)=\frac{1}{\sqrt{2\pi\alpha}}e^{i\left[\phi(x)-\theta(x)\right]},
	\end{align}
	where $\alpha$ is the ultraviolet length scale that is determined by
	the microscopic model.  The time-reversal operation
	($\mathcal{T}^2=-1$) in the bosonic language is defined as follows:
	$\phi\rightarrow -\phi+\frac{\pi}{2}$,
	$\theta\rightarrow\theta-\frac{\pi}{2}$, and
	$i\rightarrow-i$. This corresponds to the fermionic operation:
	$R\rightarrow L$, $L\rightarrow -R$, and $i\rightarrow -i$.
	
	On the contrary, the non-topological spinless fermion obeys the time-reversal operation ($\mathcal{T}^2=+1$). The fermionic operation is $R\rightarrow L$, $L\rightarrow R$, and $i\rightarrow -i$. The operation does not forbid the conventional backscattering term , $R^{\dagger}L+L^{\dagger}R$, which admits Anderson localization.
	
	\section{Rashba spin-orbit coupling and TR breaking order parameters}
	
	The TR operation $\mathcal{T}^2=-1$ forces `masses' in the helical Luttinger liquid to break TR symmetry. 
	This is because the helical Luttinger liquid still contains the parent `spinful' feature in the bulk (even though spin is not necessarily a good quantum number). We briefly discuss the connection between physical fermion and the edge chiral fermions (right and left movers).
	
	In a generic 2D topological insulator, the physical electron field can be expanded as \cite{Schmidt2012,Xie2016}
	\begin{align}
	\nonumber c_{\uparrow}(x)\approx &e^{ik_Fx}R(x)-i\zeta e^{-ik_Fx}\partial_xL(x),\\
	c_{\downarrow}(x)\approx &e^{-ik_Fx}L(x)-i\zeta^* e^{ik_Fx}\partial_xR(x),
	\end{align}
	where $\zeta$ is a phenomenological parameter that encodes the Rashba spin orbit coupling. The TR operation: $c_{\uparrow}\rightarrow c_{\downarrow}$, $c_{\downarrow}\rightarrow -c_{\uparrow}$, and $i\rightarrow -i$.
	The quantum spin Hall insulator corresponds to $\zeta=0$ due to the absence of Rashba spin orbit coupling. The electron density operator is expressed as
	\begin{align}
	\nonumber n=&c^{\dagger}_{\uparrow}c_{\uparrow}+c^{\dagger}_{\downarrow}c_{\downarrow}\\
	\approx&R^{\dagger}R+L^{\dagger}L
	-\left\{
	i\zeta e^{-i2k_Fx}\left[R^{\dagger}\partial_xL-\left(\partial_xR^{\dagger}\right)L\right]+\text{H.c.}
	\right\},
	\end{align}
	where we only keep the leading $O(|\zeta|)$ terms.
	The $2k_F$ component of the density operator contains unconventional backscattering. As pointed out in Ref.~\onlinecite{Xie2016}, the effect of these backscattering term can be dealt with change of basis, and the helical edge state remains backscattering-free in the new ``rotated frame''.
	
	The spin-flip bilinears are expressed as follows:
	\begin{align}
	c^{\dagger}_{\uparrow}c_{\downarrow}\approx & e^{-i2k_Fx}R^{\dagger}L+i\zeta^*\left(\partial_xL^{\dagger}\right)L-i\zeta^*R^{\dagger}\partial_xR,\\
	c^{\dagger}_{\downarrow}c_{\uparrow}\approx & e^{i2k_Fx}L^{\dagger}R+i\zeta\left(\partial_xR^{\dagger}\right)R-i\zeta L^{\dagger}\partial_xL,
	\end{align}
	where we have omitted the $O(|\zeta|)$ terms.
	The backscattering terms $R^{\dagger}L$ and $L^{\dagger}R$ do carry TR breaking features as the spin-flip billinears. However, they are not physical spin-flip operator as long as $\zeta$ is non-zero (due to the lack of spin conservation). In bosonization, the TR breaking mass operators are expressed as follows \cite{Wu2006}:
	\begin{align}
	m_x=&\,e^{i2k_Fx}L^{\dagger}R+\text{H.c.}=\frac{1}{\pi\alpha}\sin\left[2\theta+(2k_F-Q)x\right],\\
	m_y=&\,ie^{i2k_Fx}L^{\dagger}R+\text{H.c.}=\frac{1}{\pi\alpha}\cos\left[2\theta+(2k_F-Q)x\right],
	\end{align}
	where $Q$ is the commensurate wavevector set by the microscopic lattice model. Finite expectation of $m_x$ ($m_y$) can be viewed as a {\it pseudospin} order along $x$ ($y$) direction that breaks TR symmetry.
	
	\section{Effective Action and Renormalization Group Analysis}\label{App:RG}
	
	In the presence of the Gaussian unbounded forward disorder, 
	the interaction term $\mathcal{S}_U$ becomes irrelevant. This can be simply seen in Eq.~(\ref{eq:U'}). The fluctuating factor $\eta(x)$ vanishes after the disorder average.
	
	At second order of $U$, $\mathcal{S}_{U,dis}$ given by Eq.~(\ref{Eq:S_U_dis}) is generated under RG flow. The precise functional form of the coupling constant $\Delta_U$ is not very important for this study. We follow the strategy in Ref.~\onlinecite{Kainaris2014}.	First of all, we examine the correlation of the random field $\eta(x)$ in Eq.~(\ref{eq:U'}) in a system with finite length $L$. We can access both the clean and disorder limits. The first and second cumulants of $\eta$ are as follows:
	\begin{align}
	\overline{\eta(x)}=&e^{-\frac{8K^2}{v^2}\Delta|x+L/2|}e^{-i\delta Q x},\\
	\overline{\eta(x')\eta(x)}=&e^{-\frac{8K^2}{v^2}\Delta|x+x'+2\min(x,x')+2L|}e^{-i\delta Q(x+x')},\\
	\overline{\eta^*(x')\eta(x)}=&e^{-\frac{8K^2}{v^2}\Delta|x-x'|}e^{-i\delta
		Q(x-x')},
	\end{align}
	where $\overline{\eta(x)}$ and $\overline{\eta(x')\eta(x)}$ depend on a dimensionless parameter $K^2\Delta L/v^2$. The disorder effect is negligible when $K^2\Delta L/v^2\rightarrow 0$; the clean limit is reproduced.
	In the limit $K^2\Delta L/v^2\rightarrow \infty$, the disorder is relevant. $\overline{\eta(x)}$ and $\overline{\eta(x')\eta(x)}$ vanish. This disorder limit is focused in the main text.
	
	Then, we perform cumulant expansion of the partition function, $\overline{e^{-S_U}}\approx \exp\left[-\overline{S_U}+\frac{1}{2}\left(\overline{S_U^2}-\overline{S_U}^2\right)\right]\equiv e^{-\mathcal{S}_U^{(1)}-\mathcal{S}_U^{(2)}}$.
	The linear in $U$ contribution is given by
	\begin{align}
	\mathcal{S}_U^{(1)}=\tilde{U}\sum_a\int d\tau dx\, e^{-\frac{8K^2}{v^2}\Delta|x+L/2|}\cos\left[4\theta_a-\delta Qx\right],
	\end{align}
	where $a$ is the replica index.
	In the clean limit ($K^2\Delta L/v^2\rightarrow0$), $\mathcal{S}_U^{(1)}$ reduces to the uniform umklapp term $\mathcal{S}_U$ given by Eq.~(\ref{Eq:S_U}). $\mathcal{S}_U^{(1)}$ vanishes when taking the disorder limit ($K^2\Delta L/v^2\rightarrow \infty$).  
	The second order contribution in the action is as follows:
	\begin{align}
	\nonumber\mathcal{S}_{U}^{(2)}
	=&-\frac{1}{2}\left(\frac{\tilde{U}}{2}\right)^2\sum_{a,b}\int d\tau dx \int d\tau'dx\\
	\nonumber&\times\bigg\{\overline{\eta^*(x)\eta(x)}\,e^{i4\,\left[\theta_a(\tau,x)-\theta_b(\tau',x')\right]}\\
	\nonumber&\hspace{0.6cm}+\overline{\eta^*(x)\eta^*(x)}\,e^{i4\,\left[\theta_a(\tau,x)+\theta_b(\tau',x')\right]}+\text{H.c.}
	\bigg\}\\
	&+\frac{1}{2}\overline{S_U}^2,
	\end{align}
	where $a$ and $b$ are replica indexes.
	In the clean limit ($K^2\Delta L/v^2\rightarrow0$), $\mathcal{S}_{U}^{(2)}=0$. In the disorder limit ($K^2\Delta L/v^2\rightarrow \infty$),
	\begin{align}
	\nonumber\mathcal{S}_{U}^{(2)}=&-\frac{1}{2}\left(\frac{\tilde{U}}{2}\right)^2\sum_{a,b}\int d\tau dx \int d\tau' dx'\,e^{-\frac{8K^2\Delta}{v^2}|x-x'|}\\
	\label{Eq:S_U_2}&\times\!\left\{
	e^{i\delta Q(x-x')}e^{i4\,\left[\theta_a(\tau,x)-\theta_b(\tau',x')\right]}
	+\text{H.c.}
	\right\}.
	\end{align}
	 The exponential decay in the first line of Eq.~(\ref{Eq:S_U_2}) implies that the long wavelength physics can be treated as effective white noise correlated problem, similar to the backscattering term in the Giamarchi-Schulz model \cite{Giamarchi1988,Wu2006,Xu2006}.
	In order to see this explicitly, we integrate over the relative spatial degrees of freedom, $r=x-x'$, to compute $\Delta_U$.
	\begin{align}
	\Delta_U=&\frac{1}{4}\tilde{U}^2\int_{-\infty}^{\infty} dr \,e^{-8\frac{K^2\Delta }{v^2}|r|} e^{i\delta Q r}\\
	=&\tilde{U}^2\frac{K^2\Delta /v^2}{16(K^2\Delta /v^2)^2+\delta Q^2/4}.
	\end{align}
	Based on scaling analysis we obtain the RG flow equation:
	\begin{align}
	\frac{d\Delta_U}{dl}=\left(3-8K\right)\Delta_U.
	\end{align}
	Therefore, the dirty helical Luttinger liquid is stable for $K>3/8$ in the model given by Eq.~(\ref{Eq:S}). Meanwhile, for $K < 3/8$, backscattering is a relevant perturbation. This we interpret as an instability to localization, a conclusion that is supported by analysis of the exactly solvable Luther-Emery point $K=1/4$ in the main text.

	\section{Domain Wall Solution with Disorder}\label{App:DW_dis}
	
	We first consider a sine-Gordon model in the clean limit. The action is as follows:
	\begin{align}
	\nonumber\mathcal{S}'=&\int d\tau  dx\,\frac{1}{2\pi vK}\left[\left(\partial_{\tau}\theta\right)^2+v^2\left(\partial_x\theta\right)^2\right]\\
	\label{Eq:S_inst}&+\tilde{U}\int d\tau dx\left\{\cos\left[4\theta(\tau,x)\right]+1\right\},
	\end{align}
	where $\tilde{U}$ is the interaction strength. 
	We have assumed the system is commensurate. 
	The extra constant $1$ is added in order to avoid infinity in the later calculations. 
	The equation of motion is as follows:
	\begin{align}\label{Eq:EoM}
	\frac{1}{\pi vK}\left[\partial_{\tau}^2\theta+v^2\partial_x^2\theta\right]+4\tilde{U}\sin\left[4\theta\right]=0.
	\end{align}
	Assuming no temporal fluctuation, we can construct a domain wall solution as follows:
	\begin{align}
	\label{Eq:theta_X}\theta_X(x)=&\arctan\left[-\tanh\left(2\sqrt{\pi\tilde{U}K/v}x\right)\right],
	\end{align}
	where $\theta_X(-\infty)=-\pi/4$ and
	$\theta_X(\infty)=\pi/4$. 
	The solution $\theta_X$ is approximately correct for a finite system with size $L$ when $\sqrt{\pi\tilde{U}K/v}L\gg1$.
	
	We first focus on the saddle point contribution of $\theta_X(x)$. The partition function is approximated by $Z\propto \sum_{x_0}e^{-\mathcal{S}[\theta_X(x-x_0)]}$ where $x_0$ indicates the position of the domain wall. The saddle point contribution $\mathcal{S}[\theta_X(x-x_0)]$ only weakly depends on $x_0$. We plug Eq.~(\ref{Eq:theta_X}) in Eq.~(\ref{Eq:S_inst}).
	\begin{align}
	\nonumber\mathcal{S}[\theta_X]=&\beta\int dx\left\{\frac{v\left(\partial_x\theta_X\right)^2}{2\pi K}+\tilde{U}\left[\cos\left(4\theta_X\right)\!+\!1\right]\!\right\}\\
	\nonumber=&
	2\tilde{U}\beta\int dx\left[\cos\left(4\theta_X\right)\!+\!1\right]\\
	=&2\beta\sqrt{\tilde{U}v/(\pi K)},
	\end{align}
	where we have used Eq.~(\ref{Eq:EoM}). The energy cost of having a domain wall is $\Delta E=2\sqrt{\tilde{U}v/(\pi K)}$.

	\subsection{Increased Energy Due to Disorder}

	Now, we consider adding a disorder action $\mathcal{S}_V$ [given by (\ref{Eq:S_V})]. We first perform a change of variable to eliminate the $\mathcal{S}_V$, $\theta(x)\rightarrow
	\theta(x)-\frac{K}{v}\int_{-L/2}^{x}ds V(s)$. The $\mathcal{S}'$ becomes
	\begin{align}
	\nonumber\mathcal{S}'=&\int d\tau dx\,\frac{1}{2v\pi K}\left[\left(\partial_{\tau}\theta\right)^2+v^2\left(\partial_x\theta\right)^2\right]\\
	&+\tilde{U}\int d\tau dx\left\{\cos\left[4\tilde{\theta}(\tau,x)\right]+1\right\},
	\end{align}
	where $\tilde{\theta}(\tau,x)=\theta(\tau,x)-\frac{K}{v}\int_{-L/2}^{x}ds V(s)$.

	First of all, we assume that the field $\tilde\theta$ is pinned by the interaction term. This means $\theta(\tau,x)=\frac{K}{v}\int_{-L/2}^{x}ds V(s)+\frac{\pi}{4}(2N+1)$ with integer $N$. The solution minimizes the $\tilde{U}$ term to zero. We compute the corresponding energy in the kinetic energy part
	\begin{align}
	\delta E=&\frac{v}{2\pi K}\int dx\left(\partial_x\theta\right)^2=\frac{v}{2\pi K}\frac{K^2}{v^2}\int dx V^2(x),\\
	\rightarrow \overline{\delta E}=&\frac{K}{2v\pi}\Delta L,
	\end{align}
	where $\overline{\delta E}$ is the averaged kinetic energy. The extra energy cost of a uniform configuration due to disorder is proportional to the system size which is even stronger than in the Imry-Ma argument \cite{Imry1975}. Meanwhile, the energy cost of a domain wall is just a constant, as discussed above. Therefore, it is energetically favorable to locally distort to match the disorder, thereby introducing a finite density of domain walls even at zero temperature. 
	
	\subsection{Domain Wall Pinning in A Fixed Realization}
	We previously showed that the ground state should contain a non-zero density of domain walls. In this appendix we construct the domain wall solution in a fixed realization of disorder.

	The disorder potential $V(x)$ couples to $\partial_x\theta$. The domain wall solution is mostly flat except for the ``thin wall'' region. A domain wall can take place as long as the local disorder potential provides enough energy gain. This is translated into
	\begin{align}\label{Eq:DW_pinning}
	\left|\int dx \frac{V(x)}{\pi}\partial_x\theta_X(x-x_0)\right|\ge 2\sqrt{\tilde{U}v/(\pi K)},
	\end{align}
	where $x_0$ is the position of a domain wall and $\theta_X$ is given by Eq.~(\ref{Eq:theta_X}). The sign of the integral determines if $\theta$ should increase or decrease in value (see Fig.~\ref{Fig:DW}). Base on Eq.~(\ref{Eq:DW_pinning}), the domain walls are pinned by the rare strong disorder regions.
	The ground state consists of multiple mini blocks that are separated by domain walls. Inside each block, $\theta=(2N+1)\pi/4$ where $N$ is an integer. 
	This line of reasoning predicts a ground state that breaks TR symmetry in `spin glass' like fashion, with localized domain walls separating domains with different values of $\theta$. 
	
	\section{Luther-Emery fermion}\label{App:LE_mapping}
	At $K=1/4$, equation~(\ref{Eq:S}) can be mapped to a non-interacting Luther-Emery fermion. The Luther-Emery fermionic fields \cite{Giamarchi_Book} can be described
	by rescaled chiral bosons via
	\begin{align}
	\Psi_R(x)=\frac{1}{\sqrt{2\pi\alpha}}e^{i\left[\phi(x)/2+2\theta(x)\right]},\,\,\Psi_L(x)=\frac{1}{\sqrt{2\pi\alpha}}e^{i\left[\phi(x)/2-2\theta(x)\right]},
	\end{align}
	where $\alpha$ is the ultraviolet length scale. The Luther-Emery fermion fields are related to the physical fermions upon $\phi/2\rightarrow\phi$ and $2\theta\rightarrow \theta$.
	The TR operation $(\mathcal{T}^2=-1)$ in terms of the Luther-Emery fermions is given by $\Psi_R\rightarrow e^{i3\pi/4}\Psi_L$, $\Psi_L\rightarrow e^{i3\pi/4}\Psi_R$, and $i\rightarrow -i$. One can use the above rule to confirm the TR invariance in Eq.~(\ref{Eq:LE_H}).
	The non-trivial phase factor $e^{i3\pi/4}$ implies that the Luther-Emery fermion is non-local in terms of the physical fermionic fields $R$ and $L$. As we explained in the main text, the density of a Luther-Emery fermion is related to a domain wall, $\theta$-kink. The precise operator correspondence depends on both $\theta$ and $\phi$ bosonic fields.
	
	When $K$ is slightly tuned away from $1/4$, an interaction term arises \cite{Foster2010} and is given by
	\begin{align}
	\hat{H}_{LE,I}=\frac{v}{2K^2}\left(\frac{1}{16}-K^2\right)\left[:\Psi^{\dagger}_R\Psi_R+\Psi^{\dagger}_L\Psi_L:\right]^2.
	\end{align}
	The interaction is repulsive (attractive) for $K<1/4$ ($K>1/4$). This suggests that the refermionized theory is no solvable when $K\neq 1/4$. Meanwhile, the localized state is still stable for a range of $K\neq 1/4$ that we discussed below. 
	
	For repulsive interactions ($K<1/4$), the backscattering is enhanced \cite{Kane1992PRL,Kane1992PRB,Matveev1993,Garst2008} and the localized state remains stable. For attractive interactions ($K>1/4$), the localization becomes less stable. Even though we can not systematically determine the critical value from the Luther-Emery theory, the complementary bosonization analysis gives a critical point at $K=3/8$ which we have discussed extensively in the main text. 
	
	\section{Effects of magnetic field}\label{App:B_field}
	\label{sec:field}
	The magnetic field coupling terms at the linear order \cite{Qi2008,Maciejko2010} are as follows:
	\begin{align}
	\nonumber H_B=&t_zB_z\int dx \left[R^{\dagger}R-L^{\dagger}L\right]\\
	\nonumber&+t_xB_x\int dx \left[e^{-i2k_Fx}R^{\dagger}L+e^{i2k_Fx}L^{\dagger}R\right]\\
	\label{Eq:H_B}&+t_yB_y\int dx \left[-ie^{-i2k_Fx}R^{\dagger}L+ie^{i2k_Fx}L^{\dagger}R\right],
	\end{align}
	where $B_a$ is the $a-$component of the magnetic field, $t_a$ is the model-dependent coupling constant for the $a-$component. The $t_z$ term shifts the wavevector of the edge mode but does not induce backscattering for the weak field. For simplicity, we only consider the $t_x$ and $t_y$ terms that induce backscatterings of the edge states. 
	The corresponding bosonized action is given by,
	\begin{align}
	\nonumber\mathcal{S}_B=&\frac{t_xB_x}{\pi\alpha}\int d\tau dx \sin\left[2\theta(\tau,x)+(2k_F-Q)x\right]\\
	\label{Eq:S_B}&+\frac{t_yB_y}{\pi\alpha}\int d\tau dx \cos\left[2\theta(\tau,x)+(2k_F-Q)x\right],
	\end{align}
	where $Q$ is the commensurate wavevector set by the microscopic lattice model. 
	
	In a clean helical edge with $K>1/2$, the magnetic field term $\mathcal{S}_B$ can open up a gap when $Q=2k_F$. In particular, the conductance in a non-interacting edge state ($K=1$) drops from $G=G_0=e^2/h$ to $G(B)\approx G_0 e^{-m(B)L/v}$, where $L$ is edge length and $m(B)\propto B$ is the gap induced by the magnetic field. 
	On the other hand, the transport behavior does not change much in a clean edge with $K<1/2$ and $Q=2k_F$ (which implies $4k_F$ is also commensurate). 
	The ground state configuration is determine by minimizing $\mathcal{S}_U+\mathcal{S}_B$ rather than $\mathcal{S}_U$ alone. 
	The effect of the magnetic field is to enhance the gap and change the precise ground state condition.

	In the disordered case, we first examine the non-interacting limit ($K=1$ and $U=0$).
	The model corresponds to a massive Dirac fermion with a scalar potential disorder. This Dirac model gives Anderson localization for the whole spectrum \cite{Bocquet1999}. 
	The mass term is proportional to the strength of the magnetic field $B$. The value of $(2k_F-Q)$ in $\mathcal{S}_B$ determines the position of the fermi energy which does not affect the localization.

	For a generic interacting edge state with disorder, we adopt the same procedure of deriving $\mathcal{S}_{U,dis}$ in Sec.~\ref{Sec:QuSG}. The new effective action is as follows:
	\begin{align}
	\nonumber\mathcal{S}_{B,dis}=&-\Delta_B\sum_{a,b}\int d\tau d\tau' dx \\
	\label{Eq:S_B_dis}&\times\cos\left[2\left(\theta_a(\tau,x)-\theta_b(\tau',x)\right)\right],
	\end{align}
	where $\Delta_B\propto B^2$. $\mathcal{S}_{B,dis}$ is the same as the disorder averaged backscattering term in the Giamarchi-Schulz model \cite{Giamarchi1988}. 
	$\mathcal{S}_{B,dis}$ is relevant for $K<3/2$ and is the leading perturbation in the model. 
	In the RG analysis, the field-dependent localization length can be obtained, $l_B\propto B^{-2/(3-2K)}$.
	
%%\bibliography{TI_ref}

\begin{thebibliography}{75}%
	\makeatletter
	\providecommand \@ifxundefined [1]{%
		\@ifx{#1\undefined}
	}%
	\providecommand \@ifnum [1]{%
		\ifnum #1\expandafter \@firstoftwo
		\else \expandafter \@secondoftwo
		\fi
	}%
	\providecommand \@ifx [1]{%
		\ifx #1\expandafter \@firstoftwo
		\else \expandafter \@secondoftwo
		\fi
	}%
	\providecommand \natexlab [1]{#1}%
	\providecommand \enquote  [1]{``#1''}%
	\providecommand \bibnamefont  [1]{#1}%
	\providecommand \bibfnamefont [1]{#1}%
	\providecommand \citenamefont [1]{#1}%
	\providecommand \href@noop [0]{\@secondoftwo}%
	\providecommand \href [0]{\begingroup \@sanitize@url \@href}%
	\providecommand \@href[1]{\@@startlink{#1}\@@href}%
	\providecommand \@@href[1]{\endgroup#1\@@endlink}%
	\providecommand \@sanitize@url [0]{\catcode `\\12\catcode `\$12\catcode
		`\&12\catcode `\#12\catcode `\^12\catcode `\_12\catcode `\%12\relax}%
	\providecommand \@@startlink[1]{}%
	\providecommand \@@endlink[0]{}%
	\providecommand \url  [0]{\begingroup\@sanitize@url \@url }%
	\providecommand \@url [1]{\endgroup\@href {#1}{\urlprefix }}%
	\providecommand \urlprefix  [0]{URL }%
	\providecommand \Eprint [0]{\href }%
	\providecommand \doibase [0]{http://dx.doi.org/}%
	\providecommand \selectlanguage [0]{\@gobble}%
	\providecommand \bibinfo  [0]{\@secondoftwo}%
	\providecommand \bibfield  [0]{\@secondoftwo}%
	\providecommand \translation [1]{[#1]}%
	\providecommand \BibitemOpen [0]{}%
	\providecommand \bibitemStop [0]{}%
	\providecommand \bibitemNoStop [0]{.\EOS\space}%
	\providecommand \EOS [0]{\spacefactor3000\relax}%
	\providecommand \BibitemShut  [1]{\csname bibitem#1\endcsname}%
	\let\auto@bib@innerbib\@empty
	%</preamble>
	\bibitem [{\citenamefont {Kane}\ and\ \citenamefont
		{Mele}(2005{\natexlab{a}})}]{Kane2005_1}%
	\BibitemOpen
	\bibfield  {author} {\bibinfo {author} {\bibfnamefont {C.~L.}\ \bibnamefont
			{Kane}}\ and\ \bibinfo {author} {\bibfnamefont {E.~J.}\ \bibnamefont
			{Mele}},\ }\href@noop {} {\bibfield  {journal} {\bibinfo  {journal} {Phy.
				Rev. Lett.}\ }\textbf {\bibinfo {volume} {95}},\ \bibinfo {pages} {146802}
		(\bibinfo {year} {2005}{\natexlab{a}})}\BibitemShut {NoStop}%
	\bibitem [{\citenamefont {Hasan}\ and\ \citenamefont
		{Kane}(2010)}]{Hasan2010_RMP}%
	\BibitemOpen
	\bibfield  {author} {\bibinfo {author} {\bibfnamefont {M.~Z.}\ \bibnamefont
			{Hasan}}\ and\ \bibinfo {author} {\bibfnamefont {C.~L.}\ \bibnamefont
			{Kane}},\ }\href {\doibase 10.1103/RevModPhys.82.3045} {\bibfield  {journal}
		{\bibinfo  {journal} {Rev. Mod. Phys.}\ }\textbf {\bibinfo {volume} {82}},\
		\bibinfo {pages} {3045} (\bibinfo {year} {2010})}\BibitemShut {NoStop}%
	\bibitem [{\citenamefont {Senthil}(2015)}]{SenthilARCMP}%
	\BibitemOpen
	\bibfield  {author} {\bibinfo {author} {\bibfnamefont {T.}~\bibnamefont
			{Senthil}},\ }\href {\doibase 10.1146/annurev-conmatphys-031214-014740}
	{\bibfield  {journal} {\bibinfo  {journal} {Annual Review of Condensed Matter
				Physics}\ }\textbf {\bibinfo {volume} {6}},\ \bibinfo {pages} {299} (\bibinfo
		{year} {2015})},\ \Eprint
	{http://arxiv.org/abs/https://doi.org/10.1146/annurev-conmatphys-031214-014740}
	{https://doi.org/10.1146/annurev-conmatphys-031214-014740} \BibitemShut
	{NoStop}%
	\bibitem [{\citenamefont {Kane}\ and\ \citenamefont
		{Mele}(2005{\natexlab{b}})}]{Kane2005_2}%
	\BibitemOpen
	\bibfield  {author} {\bibinfo {author} {\bibfnamefont {C.~L.}\ \bibnamefont
			{Kane}}\ and\ \bibinfo {author} {\bibfnamefont {E.~J.}\ \bibnamefont
			{Mele}},\ }\href@noop {} {\bibfield  {journal} {\bibinfo  {journal} {Phy.
				Rev. Lett.}\ }\textbf {\bibinfo {volume} {95}},\ \bibinfo {pages} {226801}
		(\bibinfo {year} {2005}{\natexlab{b}})}\BibitemShut {NoStop}%
	\bibitem [{\citenamefont {Bernevig}\ and\ \citenamefont
		{Zhang}(2006)}]{Bernevig2006}%
	\BibitemOpen
	\bibfield  {author} {\bibinfo {author} {\bibfnamefont {B.~A.}\ \bibnamefont
			{Bernevig}}\ and\ \bibinfo {author} {\bibfnamefont {S.-C.}\ \bibnamefont
			{Zhang}},\ }\href {\doibase 10.1103/PhysRevLett.96.106802} {\bibfield
		{journal} {\bibinfo  {journal} {Phys. Rev. Lett.}\ }\textbf {\bibinfo
			{volume} {96}},\ \bibinfo {pages} {106802} (\bibinfo {year}
		{2006})}\BibitemShut {NoStop}%
	\bibitem [{\citenamefont {Fu}\ and\ \citenamefont {Kane}(2009)}]{Fu2009}%
	\BibitemOpen
	\bibfield  {author} {\bibinfo {author} {\bibfnamefont {L.}~\bibnamefont
			{Fu}}\ and\ \bibinfo {author} {\bibfnamefont {C.~L.}\ \bibnamefont {Kane}},\
	}\href {\doibase 10.1103/PhysRevB.79.161408} {\bibfield  {journal} {\bibinfo
			{journal} {Phys. Rev. B}\ }\textbf {\bibinfo {volume} {79}},\ \bibinfo
		{pages} {161408} (\bibinfo {year} {2009})}\BibitemShut {NoStop}%
	\bibitem [{\citenamefont {Zhang}\ and\ \citenamefont {Kane}(2014)}]{Zhang2014}%
	\BibitemOpen
	\bibfield  {author} {\bibinfo {author} {\bibfnamefont {F.}~\bibnamefont
			{Zhang}}\ and\ \bibinfo {author} {\bibfnamefont {C.~L.}\ \bibnamefont
			{Kane}},\ }\href {\doibase 10.1103/PhysRevLett.113.036401} {\bibfield
		{journal} {\bibinfo  {journal} {Phys. Rev. Lett.}\ }\textbf {\bibinfo
			{volume} {113}},\ \bibinfo {pages} {036401} (\bibinfo {year}
		{2014})}\BibitemShut {NoStop}%
	\bibitem [{\citenamefont {Orth}\ \emph {et~al.}(2015)\citenamefont {Orth},
		\citenamefont {Tiwari}, \citenamefont {Meng},\ and\ \citenamefont
		{Schmidt}}]{Orth2015}%
	\BibitemOpen
	\bibfield  {author} {\bibinfo {author} {\bibfnamefont {C.~P.}\ \bibnamefont
			{Orth}}, \bibinfo {author} {\bibfnamefont {R.~P.}\ \bibnamefont {Tiwari}},
		\bibinfo {author} {\bibfnamefont {T.}~\bibnamefont {Meng}}, \ and\ \bibinfo
		{author} {\bibfnamefont {T.~L.}\ \bibnamefont {Schmidt}},\ }\href {\doibase
		10.1103/PhysRevB.91.081406} {\bibfield  {journal} {\bibinfo  {journal} {Phys.
				Rev. B}\ }\textbf {\bibinfo {volume} {91}},\ \bibinfo {pages} {081406}
		(\bibinfo {year} {2015})}\BibitemShut {NoStop}%
	\bibitem [{\citenamefont {Alicea}\ and\ \citenamefont
		{Fendley}(2016)}]{Alicea2016Review}%
	\BibitemOpen
	\bibfield  {author} {\bibinfo {author} {\bibfnamefont {J.}~\bibnamefont
			{Alicea}}\ and\ \bibinfo {author} {\bibfnamefont {P.}~\bibnamefont
			{Fendley}},\ }\href@noop {} {\bibfield  {journal} {\bibinfo  {journal}
			{Annual Review of Condensed Matter Physics}\ }\textbf {\bibinfo {volume}
			{7}},\ \bibinfo {pages} {119} (\bibinfo {year} {2016})}\BibitemShut {NoStop}%
	\bibitem [{\citenamefont {Fisher}\ \emph {et~al.}(1989)\citenamefont {Fisher},
		\citenamefont {Weichman}, \citenamefont {Grinstein},\ and\ \citenamefont
		{Fisher}}]{Fisher1989}%
	\BibitemOpen
	\bibfield  {author} {\bibinfo {author} {\bibfnamefont {M.~P.~A.}\
			\bibnamefont {Fisher}}, \bibinfo {author} {\bibfnamefont {P.~B.}\
			\bibnamefont {Weichman}}, \bibinfo {author} {\bibfnamefont {G.}~\bibnamefont
			{Grinstein}}, \ and\ \bibinfo {author} {\bibfnamefont {D.~S.}\ \bibnamefont
			{Fisher}},\ }\href {\doibase 10.1103/PhysRevB.40.546} {\bibfield  {journal}
		{\bibinfo  {journal} {Phys. Rev. B}\ }\textbf {\bibinfo {volume} {40}},\
		\bibinfo {pages} {546} (\bibinfo {year} {1989})}\BibitemShut {NoStop}%
	\bibitem [{\citenamefont {K{\"o}nig}\ \emph {et~al.}(2007)\citenamefont
		{K{\"o}nig}, \citenamefont {Wiedmann}, \citenamefont {Br{\"u}ne},
		\citenamefont {Roth}, \citenamefont {Buhmann}, \citenamefont {Molenkamp},
		\citenamefont {Qi},\ and\ \citenamefont {Zhang}}]{Konig2007}%
	\BibitemOpen
	\bibfield  {author} {\bibinfo {author} {\bibfnamefont {M.}~\bibnamefont
			{K{\"o}nig}}, \bibinfo {author} {\bibfnamefont {S.}~\bibnamefont {Wiedmann}},
		\bibinfo {author} {\bibfnamefont {C.}~\bibnamefont {Br{\"u}ne}}, \bibinfo
		{author} {\bibfnamefont {A.}~\bibnamefont {Roth}}, \bibinfo {author}
		{\bibfnamefont {H.}~\bibnamefont {Buhmann}}, \bibinfo {author} {\bibfnamefont
			{L.~W.}\ \bibnamefont {Molenkamp}}, \bibinfo {author} {\bibfnamefont {X.-L.}\
			\bibnamefont {Qi}}, \ and\ \bibinfo {author} {\bibfnamefont {S.-C.}\
			\bibnamefont {Zhang}},\ }\href@noop {} {\bibfield  {journal} {\bibinfo
			{journal} {Science}\ }\textbf {\bibinfo {volume} {318}},\ \bibinfo {pages}
		{766} (\bibinfo {year} {2007})}\BibitemShut {NoStop}%
	\bibitem [{\citenamefont {Knez}\ \emph {et~al.}(2011)\citenamefont {Knez},
		\citenamefont {Du},\ and\ \citenamefont {Sullivan}}]{Knez2011}%
	\BibitemOpen
	\bibfield  {author} {\bibinfo {author} {\bibfnamefont {I.}~\bibnamefont
			{Knez}}, \bibinfo {author} {\bibfnamefont {R.-R.}\ \bibnamefont {Du}}, \ and\
		\bibinfo {author} {\bibfnamefont {G.}~\bibnamefont {Sullivan}},\ }\href
	{\doibase 10.1103/PhysRevLett.107.136603} {\bibfield  {journal} {\bibinfo
			{journal} {Phys. Rev. Lett.}\ }\textbf {\bibinfo {volume} {107}},\ \bibinfo
		{pages} {136603} (\bibinfo {year} {2011})}\BibitemShut {NoStop}%
	\bibitem [{\citenamefont {Suzuki}\ \emph {et~al.}(2013)\citenamefont {Suzuki},
		\citenamefont {Harada}, \citenamefont {Onomitsu},\ and\ \citenamefont
		{Muraki}}]{Suzuki2013}%
	\BibitemOpen
	\bibfield  {author} {\bibinfo {author} {\bibfnamefont {K.}~\bibnamefont
			{Suzuki}}, \bibinfo {author} {\bibfnamefont {Y.}~\bibnamefont {Harada}},
		\bibinfo {author} {\bibfnamefont {K.}~\bibnamefont {Onomitsu}}, \ and\
		\bibinfo {author} {\bibfnamefont {K.}~\bibnamefont {Muraki}},\ }\href
	{\doibase 10.1103/PhysRevB.87.235311} {\bibfield  {journal} {\bibinfo
			{journal} {Phys. Rev. B}\ }\textbf {\bibinfo {volume} {87}},\ \bibinfo
		{pages} {235311} (\bibinfo {year} {2013})}\BibitemShut {NoStop}%
	\bibitem [{\citenamefont {Du}\ \emph {et~al.}(2015)\citenamefont {Du},
		\citenamefont {Knez}, \citenamefont {Sullivan},\ and\ \citenamefont
		{Du}}]{Du2015}%
	\BibitemOpen
	\bibfield  {author} {\bibinfo {author} {\bibfnamefont {L.}~\bibnamefont
			{Du}}, \bibinfo {author} {\bibfnamefont {I.}~\bibnamefont {Knez}}, \bibinfo
		{author} {\bibfnamefont {G.}~\bibnamefont {Sullivan}}, \ and\ \bibinfo
		{author} {\bibfnamefont {R.-R.}\ \bibnamefont {Du}},\ }\href {\doibase
		10.1103/PhysRevLett.114.096802} {\bibfield  {journal} {\bibinfo  {journal}
			{Phys. Rev. Lett.}\ }\textbf {\bibinfo {volume} {114}},\ \bibinfo {pages}
		{096802} (\bibinfo {year} {2015})}\BibitemShut {NoStop}%
	\bibitem [{\citenamefont {Li}\ \emph {et~al.}(2015)\citenamefont {Li},
		\citenamefont {Wang}, \citenamefont {Fu}, \citenamefont {Du}, \citenamefont
		{Schreiber}, \citenamefont {Mu}, \citenamefont {Liu}, \citenamefont
		{Sullivan}, \citenamefont {Cs\'athy}, \citenamefont {Lin},\ and\
		\citenamefont {Du}}]{Li2015}%
	\BibitemOpen
	\bibfield  {author} {\bibinfo {author} {\bibfnamefont {T.}~\bibnamefont
			{Li}}, \bibinfo {author} {\bibfnamefont {P.}~\bibnamefont {Wang}}, \bibinfo
		{author} {\bibfnamefont {H.}~\bibnamefont {Fu}}, \bibinfo {author}
		{\bibfnamefont {L.}~\bibnamefont {Du}}, \bibinfo {author} {\bibfnamefont
			{K.~A.}\ \bibnamefont {Schreiber}}, \bibinfo {author} {\bibfnamefont
			{X.}~\bibnamefont {Mu}}, \bibinfo {author} {\bibfnamefont {X.}~\bibnamefont
			{Liu}}, \bibinfo {author} {\bibfnamefont {G.}~\bibnamefont {Sullivan}},
		\bibinfo {author} {\bibfnamefont {G.~A.}\ \bibnamefont {Cs\'athy}}, \bibinfo
		{author} {\bibfnamefont {X.}~\bibnamefont {Lin}}, \ and\ \bibinfo {author}
		{\bibfnamefont {R.-R.}\ \bibnamefont {Du}},\ }\href {\doibase
		10.1103/PhysRevLett.115.136804} {\bibfield  {journal} {\bibinfo  {journal}
			{Phys. Rev. Lett.}\ }\textbf {\bibinfo {volume} {115}},\ \bibinfo {pages}
		{136804} (\bibinfo {year} {2015})}\BibitemShut {NoStop}%
	\bibitem [{\citenamefont {Qu}\ \emph {et~al.}(2015)\citenamefont {Qu},
		\citenamefont {Beukman}, \citenamefont {Nadj-Perge}, \citenamefont {Wimmer},
		\citenamefont {Nguyen}, \citenamefont {Yi}, \citenamefont {Thorp},
		\citenamefont {Sokolich}, \citenamefont {Kiselev}, \citenamefont {Manfra},
		\citenamefont {Marcus},\ and\ \citenamefont {Kouwenhoven}}]{Qu2015}%
	\BibitemOpen
	\bibfield  {author} {\bibinfo {author} {\bibfnamefont {F.}~\bibnamefont
			{Qu}}, \bibinfo {author} {\bibfnamefont {A.~J.~A.}\ \bibnamefont {Beukman}},
		\bibinfo {author} {\bibfnamefont {S.}~\bibnamefont {Nadj-Perge}}, \bibinfo
		{author} {\bibfnamefont {M.}~\bibnamefont {Wimmer}}, \bibinfo {author}
		{\bibfnamefont {B.-M.}\ \bibnamefont {Nguyen}}, \bibinfo {author}
		{\bibfnamefont {W.}~\bibnamefont {Yi}}, \bibinfo {author} {\bibfnamefont
			{J.}~\bibnamefont {Thorp}}, \bibinfo {author} {\bibfnamefont
			{M.}~\bibnamefont {Sokolich}}, \bibinfo {author} {\bibfnamefont {A.~A.}\
			\bibnamefont {Kiselev}}, \bibinfo {author} {\bibfnamefont {M.~J.}\
			\bibnamefont {Manfra}}, \bibinfo {author} {\bibfnamefont {C.~M.}\
			\bibnamefont {Marcus}}, \ and\ \bibinfo {author} {\bibfnamefont {L.~P.}\
			\bibnamefont {Kouwenhoven}},\ }\href {\doibase
		10.1103/PhysRevLett.115.036803} {\bibfield  {journal} {\bibinfo  {journal}
			{Phys. Rev. Lett.}\ }\textbf {\bibinfo {volume} {115}},\ \bibinfo {pages}
		{036803} (\bibinfo {year} {2015})}\BibitemShut {NoStop}%
	\bibitem [{\citenamefont {Nichele}\ \emph {et~al.}(2016)\citenamefont
		{Nichele}, \citenamefont {Suominen}, \citenamefont {Kjaergaard},
		\citenamefont {Marcus}, \citenamefont {Sajadi}, \citenamefont {Folk},
		\citenamefont {Qu}, \citenamefont {Beukman}, \citenamefont {de~Vries},
		\citenamefont {van Veen} \emph {et~al.}}]{Nichele2016}%
	\BibitemOpen
	\bibfield  {author} {\bibinfo {author} {\bibfnamefont {F.}~\bibnamefont
			{Nichele}}, \bibinfo {author} {\bibfnamefont {H.~J.}\ \bibnamefont
			{Suominen}}, \bibinfo {author} {\bibfnamefont {M.}~\bibnamefont
			{Kjaergaard}}, \bibinfo {author} {\bibfnamefont {C.~M.}\ \bibnamefont
			{Marcus}}, \bibinfo {author} {\bibfnamefont {E.}~\bibnamefont {Sajadi}},
		\bibinfo {author} {\bibfnamefont {J.~A.}\ \bibnamefont {Folk}}, \bibinfo
		{author} {\bibfnamefont {F.}~\bibnamefont {Qu}}, \bibinfo {author}
		{\bibfnamefont {A.~J.}\ \bibnamefont {Beukman}}, \bibinfo {author}
		{\bibfnamefont {F.~K.}\ \bibnamefont {de~Vries}}, \bibinfo {author}
		{\bibfnamefont {J.}~\bibnamefont {van Veen}},  \emph {et~al.},\ }\href@noop
	{} {\bibfield  {journal} {\bibinfo  {journal} {New Journal of Physics}\
		}\textbf {\bibinfo {volume} {18}},\ \bibinfo {pages} {083005} (\bibinfo
		{year} {2016})}\BibitemShut {NoStop}%
	\bibitem [{\citenamefont {Nguyen}\ \emph {et~al.}(2016)\citenamefont {Nguyen},
		\citenamefont {Kiselev}, \citenamefont {Noah}, \citenamefont {Yi},
		\citenamefont {Qu}, \citenamefont {Beukman}, \citenamefont {de~Vries},
		\citenamefont {van Veen}, \citenamefont {Nadj-Perge}, \citenamefont
		{Kouwenhoven}, \citenamefont {Kjaergaard}, \citenamefont {Suominen},
		\citenamefont {Nichele}, \citenamefont {Marcus}, \citenamefont {Manfra},\
		and\ \citenamefont {Sokolich}}]{Nguyen2016}%
	\BibitemOpen
	\bibfield  {author} {\bibinfo {author} {\bibfnamefont {B.-M.}\ \bibnamefont
			{Nguyen}}, \bibinfo {author} {\bibfnamefont {A.~A.}\ \bibnamefont {Kiselev}},
		\bibinfo {author} {\bibfnamefont {R.}~\bibnamefont {Noah}}, \bibinfo {author}
		{\bibfnamefont {W.}~\bibnamefont {Yi}}, \bibinfo {author} {\bibfnamefont
			{F.}~\bibnamefont {Qu}}, \bibinfo {author} {\bibfnamefont {A.~J.~A.}\
			\bibnamefont {Beukman}}, \bibinfo {author} {\bibfnamefont {F.~K.}\
			\bibnamefont {de~Vries}}, \bibinfo {author} {\bibfnamefont {J.}~\bibnamefont
			{van Veen}}, \bibinfo {author} {\bibfnamefont {S.}~\bibnamefont
			{Nadj-Perge}}, \bibinfo {author} {\bibfnamefont {L.~P.}\ \bibnamefont
			{Kouwenhoven}}, \bibinfo {author} {\bibfnamefont {M.}~\bibnamefont
			{Kjaergaard}}, \bibinfo {author} {\bibfnamefont {H.~J.}\ \bibnamefont
			{Suominen}}, \bibinfo {author} {\bibfnamefont {F.}~\bibnamefont {Nichele}},
		\bibinfo {author} {\bibfnamefont {C.~M.}\ \bibnamefont {Marcus}}, \bibinfo
		{author} {\bibfnamefont {M.~J.}\ \bibnamefont {Manfra}}, \ and\ \bibinfo
		{author} {\bibfnamefont {M.}~\bibnamefont {Sokolich}},\ }\href {\doibase
		10.1103/PhysRevLett.117.077701} {\bibfield  {journal} {\bibinfo  {journal}
			{Phys. Rev. Lett.}\ }\textbf {\bibinfo {volume} {117}},\ \bibinfo {pages}
		{077701} (\bibinfo {year} {2016})}\BibitemShut {NoStop}%
	\bibitem [{\citenamefont {Cou\"edo}\ \emph {et~al.}(2016)\citenamefont
		{Cou\"edo}, \citenamefont {Irie}, \citenamefont {Suzuki}, \citenamefont
		{Onomitsu},\ and\ \citenamefont {Muraki}}]{Couedo2016}%
	\BibitemOpen
	\bibfield  {author} {\bibinfo {author} {\bibfnamefont {F.}~\bibnamefont
			{Cou\"edo}}, \bibinfo {author} {\bibfnamefont {H.}~\bibnamefont {Irie}},
		\bibinfo {author} {\bibfnamefont {K.}~\bibnamefont {Suzuki}}, \bibinfo
		{author} {\bibfnamefont {K.}~\bibnamefont {Onomitsu}}, \ and\ \bibinfo
		{author} {\bibfnamefont {K.}~\bibnamefont {Muraki}},\ }\href {\doibase
		10.1103/PhysRevB.94.035301} {\bibfield  {journal} {\bibinfo  {journal} {Phys.
				Rev. B}\ }\textbf {\bibinfo {volume} {94}},\ \bibinfo {pages} {035301}
		(\bibinfo {year} {2016})}\BibitemShut {NoStop}%
	\bibitem [{\citenamefont {Fei}\ \emph {et~al.}(2017)\citenamefont {Fei},
		\citenamefont {Palomaki}, \citenamefont {Wu}, \citenamefont {Zhao},
		\citenamefont {Cai}, \citenamefont {Sun}, \citenamefont {Nguyen},
		\citenamefont {Finney}, \citenamefont {Xu},\ and\ \citenamefont
		{Cobden}}]{Fei2017}%
	\BibitemOpen
	\bibfield  {author} {\bibinfo {author} {\bibfnamefont {Z.}~\bibnamefont
			{Fei}}, \bibinfo {author} {\bibfnamefont {T.}~\bibnamefont {Palomaki}},
		\bibinfo {author} {\bibfnamefont {S.}~\bibnamefont {Wu}}, \bibinfo {author}
		{\bibfnamefont {W.}~\bibnamefont {Zhao}}, \bibinfo {author} {\bibfnamefont
			{X.}~\bibnamefont {Cai}}, \bibinfo {author} {\bibfnamefont {B.}~\bibnamefont
			{Sun}}, \bibinfo {author} {\bibfnamefont {P.}~\bibnamefont {Nguyen}},
		\bibinfo {author} {\bibfnamefont {J.}~\bibnamefont {Finney}}, \bibinfo
		{author} {\bibfnamefont {X.}~\bibnamefont {Xu}}, \ and\ \bibinfo {author}
		{\bibfnamefont {D.~H.}\ \bibnamefont {Cobden}},\ }\href@noop {} {\bibfield
		{journal} {\bibinfo  {journal} {Nature Physics}\ }\textbf {\bibinfo {volume}
			{13}},\ \bibinfo {pages} {677} (\bibinfo {year} {2017})}\BibitemShut
	{NoStop}%
	\bibitem [{\citenamefont {Du}\ \emph {et~al.}(2017)\citenamefont {Du},
		\citenamefont {Li}, \citenamefont {Lou}, \citenamefont {Wu}, \citenamefont
		{Liu}, \citenamefont {Han}, \citenamefont {Zhang}, \citenamefont {Sullivan},
		\citenamefont {Ikhlassi}, \citenamefont {Chang},\ and\ \citenamefont
		{Du}}]{Du2017}%
	\BibitemOpen
	\bibfield  {author} {\bibinfo {author} {\bibfnamefont {L.}~\bibnamefont
			{Du}}, \bibinfo {author} {\bibfnamefont {T.}~\bibnamefont {Li}}, \bibinfo
		{author} {\bibfnamefont {W.}~\bibnamefont {Lou}}, \bibinfo {author}
		{\bibfnamefont {X.}~\bibnamefont {Wu}}, \bibinfo {author} {\bibfnamefont
			{X.}~\bibnamefont {Liu}}, \bibinfo {author} {\bibfnamefont {Z.}~\bibnamefont
			{Han}}, \bibinfo {author} {\bibfnamefont {C.}~\bibnamefont {Zhang}}, \bibinfo
		{author} {\bibfnamefont {G.}~\bibnamefont {Sullivan}}, \bibinfo {author}
		{\bibfnamefont {A.}~\bibnamefont {Ikhlassi}}, \bibinfo {author}
		{\bibfnamefont {K.}~\bibnamefont {Chang}}, \ and\ \bibinfo {author}
		{\bibfnamefont {R.-R.}\ \bibnamefont {Du}},\ }\href {\doibase
		10.1103/PhysRevLett.119.056803} {\bibfield  {journal} {\bibinfo  {journal}
			{Phys. Rev. Lett.}\ }\textbf {\bibinfo {volume} {119}},\ \bibinfo {pages}
		{056803} (\bibinfo {year} {2017})}\BibitemShut {NoStop}%
	\bibitem [{\citenamefont {Ma}\ \emph {et~al.}(2015)\citenamefont {Ma},
		\citenamefont {Calvo}, \citenamefont {Wang}, \citenamefont {Lian},
		\citenamefont {M{\"u}hlbauer}, \citenamefont {Br{\"u}ne}, \citenamefont
		{Cui}, \citenamefont {Lai}, \citenamefont {Kundhikanjana}, \citenamefont
		{Yang}, \citenamefont {Baenninger}, \citenamefont {K{\"o}nig}, \citenamefont
		{Ames}, \citenamefont {Buhmann}, \citenamefont {Leubner}, \citenamefont
		{Molenkamp}, \citenamefont {Zhang}, \citenamefont {Goldhaber-Gordon},
		\citenamefont {Kelly},\ and\ \citenamefont {Shen}}]{Ma2015}%
	\BibitemOpen
	\bibfield  {author} {\bibinfo {author} {\bibfnamefont {E.~Y.}\ \bibnamefont
			{Ma}}, \bibinfo {author} {\bibfnamefont {M.~R.}\ \bibnamefont {Calvo}},
		\bibinfo {author} {\bibfnamefont {J.}~\bibnamefont {Wang}}, \bibinfo {author}
		{\bibfnamefont {B.}~\bibnamefont {Lian}}, \bibinfo {author} {\bibfnamefont
			{M.}~\bibnamefont {M{\"u}hlbauer}}, \bibinfo {author} {\bibfnamefont
			{C.}~\bibnamefont {Br{\"u}ne}}, \bibinfo {author} {\bibfnamefont {Y.-T.}\
			\bibnamefont {Cui}}, \bibinfo {author} {\bibfnamefont {K.}~\bibnamefont
			{Lai}}, \bibinfo {author} {\bibfnamefont {W.}~\bibnamefont {Kundhikanjana}},
		\bibinfo {author} {\bibfnamefont {Y.}~\bibnamefont {Yang}}, \bibinfo {author}
		{\bibfnamefont {M.}~\bibnamefont {Baenninger}}, \bibinfo {author}
		{\bibfnamefont {M.}~\bibnamefont {K{\"o}nig}}, \bibinfo {author}
		{\bibfnamefont {C.}~\bibnamefont {Ames}}, \bibinfo {author} {\bibfnamefont
			{H.}~\bibnamefont {Buhmann}}, \bibinfo {author} {\bibfnamefont
			{P.}~\bibnamefont {Leubner}}, \bibinfo {author} {\bibfnamefont {L.~W.}\
			\bibnamefont {Molenkamp}}, \bibinfo {author} {\bibfnamefont {S.-C.}\
			\bibnamefont {Zhang}}, \bibinfo {author} {\bibfnamefont {D.}~\bibnamefont
			{Goldhaber-Gordon}}, \bibinfo {author} {\bibfnamefont {M.~A.}\ \bibnamefont
			{Kelly}}, \ and\ \bibinfo {author} {\bibfnamefont {Z.-X.}\ \bibnamefont
			{Shen}},\ }\href@noop {} {\bibfield  {journal} {\bibinfo  {journal} {Nature
				communications}\ }\textbf {\bibinfo {volume} {6}},\ \bibinfo {pages} {7252}
		(\bibinfo {year} {2015})}\BibitemShut {NoStop}%
	\bibitem [{\citenamefont {Kane}\ and\ \citenamefont
		{Fisher}(1992{\natexlab{a}})}]{Kane1992PRL}%
	\BibitemOpen
	\bibfield  {author} {\bibinfo {author} {\bibfnamefont {C.~L.}\ \bibnamefont
			{Kane}}\ and\ \bibinfo {author} {\bibfnamefont {M.~P.~A.}\ \bibnamefont
			{Fisher}},\ }\href {\doibase 10.1103/PhysRevLett.68.1220} {\bibfield
		{journal} {\bibinfo  {journal} {Phys. Rev. Lett.}\ }\textbf {\bibinfo
			{volume} {68}},\ \bibinfo {pages} {1220} (\bibinfo {year}
		{1992}{\natexlab{a}})}\BibitemShut {NoStop}%
	\bibitem [{\citenamefont {Kane}\ and\ \citenamefont
		{Fisher}(1992{\natexlab{b}})}]{Kane1992PRB}%
	\BibitemOpen
	\bibfield  {author} {\bibinfo {author} {\bibfnamefont {C.~L.}\ \bibnamefont
			{Kane}}\ and\ \bibinfo {author} {\bibfnamefont {M.~P.~A.}\ \bibnamefont
			{Fisher}},\ }\href {\doibase 10.1103/PhysRevB.46.15233} {\bibfield  {journal}
		{\bibinfo  {journal} {Phys. Rev. B}\ }\textbf {\bibinfo {volume} {46}},\
		\bibinfo {pages} {15233} (\bibinfo {year} {1992}{\natexlab{b}})}\BibitemShut
	{NoStop}%
	\bibitem [{\citenamefont {Kane}\ and\ \citenamefont {Fisher}(1997)}]{Kane1997}%
	\BibitemOpen
	\bibfield  {author} {\bibinfo {author} {\bibfnamefont {C.~L.}\ \bibnamefont
			{Kane}}\ and\ \bibinfo {author} {\bibfnamefont {M.~P.~A.}\ \bibnamefont
			{Fisher}},\ }\href {\doibase 10.1103/PhysRevB.55.15832} {\bibfield  {journal}
		{\bibinfo  {journal} {Phys. Rev. B}\ }\textbf {\bibinfo {volume} {55}},\
		\bibinfo {pages} {15832} (\bibinfo {year} {1997})}\BibitemShut {NoStop}%
	\bibitem [{\citenamefont {Wu}\ \emph {et~al.}(2006)\citenamefont {Wu},
		\citenamefont {Bernevig},\ and\ \citenamefont {Zhang}}]{Wu2006}%
	\BibitemOpen
	\bibfield  {author} {\bibinfo {author} {\bibfnamefont {C.}~\bibnamefont
			{Wu}}, \bibinfo {author} {\bibfnamefont {B.~A.}\ \bibnamefont {Bernevig}}, \
		and\ \bibinfo {author} {\bibfnamefont {S.-C.}\ \bibnamefont {Zhang}},\ }\href
	{\doibase 10.1103/PhysRevLett.96.106401} {\bibfield  {journal} {\bibinfo
			{journal} {Phys. Rev. Lett.}\ }\textbf {\bibinfo {volume} {96}},\ \bibinfo
		{pages} {106401} (\bibinfo {year} {2006})}\BibitemShut {NoStop}%
	\bibitem [{\citenamefont {Xu}\ and\ \citenamefont {Moore}(2006)}]{Xu2006}%
	\BibitemOpen
	\bibfield  {author} {\bibinfo {author} {\bibfnamefont {C.}~\bibnamefont
			{Xu}}\ and\ \bibinfo {author} {\bibfnamefont {J.~E.}\ \bibnamefont {Moore}},\
	}\href {\doibase 10.1103/PhysRevB.73.045322} {\bibfield  {journal} {\bibinfo
			{journal} {Phys. Rev. B}\ }\textbf {\bibinfo {volume} {73}},\ \bibinfo
		{pages} {045322} (\bibinfo {year} {2006})}\BibitemShut {NoStop}%
	\bibitem [{\citenamefont {Xie}\ \emph {et~al.}(2016)\citenamefont {Xie},
		\citenamefont {Li}, \citenamefont {Chou},\ and\ \citenamefont
		{Foster}}]{Xie2016}%
	\BibitemOpen
	\bibfield  {author} {\bibinfo {author} {\bibfnamefont {H.-Y.}\ \bibnamefont
			{Xie}}, \bibinfo {author} {\bibfnamefont {H.}~\bibnamefont {Li}}, \bibinfo
		{author} {\bibfnamefont {Y.-Z.}\ \bibnamefont {Chou}}, \ and\ \bibinfo
		{author} {\bibfnamefont {M.~S.}\ \bibnamefont {Foster}},\ }\href {\doibase
		10.1103/PhysRevLett.116.086603} {\bibfield  {journal} {\bibinfo  {journal}
			{Phys. Rev. Lett.}\ }\textbf {\bibinfo {volume} {116}},\ \bibinfo {pages}
		{086603} (\bibinfo {year} {2016})}\BibitemShut {NoStop}%
	\bibitem [{\citenamefont {Edwards}\ and\ \citenamefont
		{Anderson}(1975)}]{Edwards1975}%
	\BibitemOpen
	\bibfield  {author} {\bibinfo {author} {\bibfnamefont {S.~F.}\ \bibnamefont
			{Edwards}}\ and\ \bibinfo {author} {\bibfnamefont {P.~W.}\ \bibnamefont
			{Anderson}},\ }\href@noop {} {\bibfield  {journal} {\bibinfo  {journal}
			{Journal of Physics F: Metal Physics}\ }\textbf {\bibinfo {volume} {5}},\
		\bibinfo {pages} {965} (\bibinfo {year} {1975})}\BibitemShut {NoStop}%
	\bibitem [{\citenamefont {Anderson}(1958)}]{pwa1958}%
	\BibitemOpen
	\bibfield  {author} {\bibinfo {author} {\bibfnamefont {P.~W.}\ \bibnamefont
			{Anderson}},\ }\href {\doibase 10.1103/PhysRev.109.1492} {\bibfield
		{journal} {\bibinfo  {journal} {Phys. Rev.}\ }\textbf {\bibinfo {volume}
			{109}},\ \bibinfo {pages} {1492} (\bibinfo {year} {1958})}\BibitemShut
	{NoStop}%
	\bibitem [{\citenamefont {Huse}\ \emph {et~al.}(2013)\citenamefont {Huse},
		\citenamefont {Nandkishore}, \citenamefont {Oganesyan}, \citenamefont {Pal},\
		and\ \citenamefont {Sondhi}}]{LPQO}%
	\BibitemOpen
	\bibfield  {author} {\bibinfo {author} {\bibfnamefont {D.~A.}\ \bibnamefont
			{Huse}}, \bibinfo {author} {\bibfnamefont {R.}~\bibnamefont {Nandkishore}},
		\bibinfo {author} {\bibfnamefont {V.}~\bibnamefont {Oganesyan}}, \bibinfo
		{author} {\bibfnamefont {A.}~\bibnamefont {Pal}}, \ and\ \bibinfo {author}
		{\bibfnamefont {S.~L.}\ \bibnamefont {Sondhi}},\ }\href {\doibase
		10.1103/PhysRevB.88.014206} {\bibfield  {journal} {\bibinfo  {journal} {Phys.
				Rev. B}\ }\textbf {\bibinfo {volume} {88}},\ \bibinfo {pages} {014206}
		(\bibinfo {year} {2013})}\BibitemShut {NoStop}%
	\bibitem [{\citenamefont {Nandkishore}\ and\ \citenamefont
		{Sondhi}(2017)}]{LRMBL}%
	\BibitemOpen
	\bibfield  {author} {\bibinfo {author} {\bibfnamefont {R.~M.}\ \bibnamefont
			{Nandkishore}}\ and\ \bibinfo {author} {\bibfnamefont {S.~L.}\ \bibnamefont
			{Sondhi}},\ }\href {\doibase 10.1103/PhysRevX.7.041021} {\bibfield  {journal}
		{\bibinfo  {journal} {Phys. Rev. X}\ }\textbf {\bibinfo {volume} {7}},\
		\bibinfo {pages} {041021} (\bibinfo {year} {2017})}\BibitemShut {NoStop}%
	\bibitem [{\citenamefont {Parameswaran}\ and\ \citenamefont
		{Gopalakrishnan}(2017)}]{Parameswaran_NF_glass}%
	\BibitemOpen
	\bibfield  {author} {\bibinfo {author} {\bibfnamefont {S.~A.}\ \bibnamefont
			{Parameswaran}}\ and\ \bibinfo {author} {\bibfnamefont {S.}~\bibnamefont
			{Gopalakrishnan}},\ }\href {\doibase 10.1103/PhysRevLett.119.146601}
	{\bibfield  {journal} {\bibinfo  {journal} {Phys. Rev. Lett.}\ }\textbf
		{\bibinfo {volume} {119}},\ \bibinfo {pages} {146601} (\bibinfo {year}
		{2017})}\BibitemShut {NoStop}%
	\bibitem [{\citenamefont {Li}\ \emph {et~al.}(2017)\citenamefont {Li},
		\citenamefont {Wang}, \citenamefont {Sullivan}, \citenamefont {Lin},\ and\
		\citenamefont {Du}}]{Li2017}%
	\BibitemOpen
	\bibfield  {author} {\bibinfo {author} {\bibfnamefont {T.}~\bibnamefont
			{Li}}, \bibinfo {author} {\bibfnamefont {P.}~\bibnamefont {Wang}}, \bibinfo
		{author} {\bibfnamefont {G.}~\bibnamefont {Sullivan}}, \bibinfo {author}
		{\bibfnamefont {X.}~\bibnamefont {Lin}}, \ and\ \bibinfo {author}
		{\bibfnamefont {R.-R.}\ \bibnamefont {Du}},\ }\href {\doibase
		10.1103/PhysRevB.96.241406} {\bibfield  {journal} {\bibinfo  {journal} {Phys.
				Rev. B}\ }\textbf {\bibinfo {volume} {96}},\ \bibinfo {pages} {241406}
		(\bibinfo {year} {2017})}\BibitemShut {NoStop}%
	\bibitem [{\citenamefont {Str\"om}\ \emph {et~al.}(2010)\citenamefont
		{Str\"om}, \citenamefont {Johannesson},\ and\ \citenamefont
		{Japaridze}}]{Strom2010}%
	\BibitemOpen
	\bibfield  {author} {\bibinfo {author} {\bibfnamefont {A.}~\bibnamefont
			{Str\"om}}, \bibinfo {author} {\bibfnamefont {H.}~\bibnamefont
			{Johannesson}}, \ and\ \bibinfo {author} {\bibfnamefont {G.~I.}\ \bibnamefont
			{Japaridze}},\ }\href {\doibase 10.1103/PhysRevLett.104.256804} {\bibfield
		{journal} {\bibinfo  {journal} {Phys. Rev. Lett.}\ }\textbf {\bibinfo
			{volume} {104}},\ \bibinfo {pages} {256804} (\bibinfo {year}
		{2010})}\BibitemShut {NoStop}%
	\bibitem [{\citenamefont {Budich}\ \emph {et~al.}(2012)\citenamefont {Budich},
		\citenamefont {Dolcini}, \citenamefont {Recher},\ and\ \citenamefont
		{Trauzettel}}]{Budich2012}%
	\BibitemOpen
	\bibfield  {author} {\bibinfo {author} {\bibfnamefont {J.~C.}\ \bibnamefont
			{Budich}}, \bibinfo {author} {\bibfnamefont {F.}~\bibnamefont {Dolcini}},
		\bibinfo {author} {\bibfnamefont {P.}~\bibnamefont {Recher}}, \ and\ \bibinfo
		{author} {\bibfnamefont {B.}~\bibnamefont {Trauzettel}},\ }\href {\doibase
		10.1103/PhysRevLett.108.086602} {\bibfield  {journal} {\bibinfo  {journal}
			{Phys. Rev. Lett.}\ }\textbf {\bibinfo {volume} {108}},\ \bibinfo {pages}
		{086602} (\bibinfo {year} {2012})}\BibitemShut {NoStop}%
	\bibitem [{\citenamefont {Schmidt}\ \emph {et~al.}(2012)\citenamefont
		{Schmidt}, \citenamefont {Rachel}, \citenamefont {von Oppen},\ and\
		\citenamefont {Glazman}}]{Schmidt2012}%
	\BibitemOpen
	\bibfield  {author} {\bibinfo {author} {\bibfnamefont {T.~L.}\ \bibnamefont
			{Schmidt}}, \bibinfo {author} {\bibfnamefont {S.}~\bibnamefont {Rachel}},
		\bibinfo {author} {\bibfnamefont {F.}~\bibnamefont {von Oppen}}, \ and\
		\bibinfo {author} {\bibfnamefont {L.~I.}\ \bibnamefont {Glazman}},\
	}\href@noop {} {\bibfield  {journal} {\bibinfo  {journal} {Phy. Rev. Lett.}\
		}\textbf {\bibinfo {volume} {108}},\ \bibinfo {pages} {156402} (\bibinfo
		{year} {2012})}\BibitemShut {NoStop}%
	\bibitem [{\citenamefont {Shankar}(1995)}]{Shankar1995}%
	\BibitemOpen
	\bibfield  {author} {\bibinfo {author} {\bibfnamefont {R.}~\bibnamefont
			{Shankar}},\ }\href@noop {} {\bibfield  {journal} {\bibinfo  {journal} {Acta
				Phys. Pol. B}\ }\textbf {\bibinfo {volume} {26}},\ \bibinfo {pages} {1835}
		(\bibinfo {year} {1995})}\BibitemShut {NoStop}%
	\bibitem [{\citenamefont {Giamarchi}(2004)}]{Giamarchi_Book}%
	\BibitemOpen
	\bibfield  {author} {\bibinfo {author} {\bibfnamefont {T.}~\bibnamefont
			{Giamarchi}},\ }\href@noop {} {\emph {\bibinfo {title} {Quantum physics in
				one dimension}}}\ (\bibinfo  {publisher} {Oxford Science Publications},\
	\bibinfo {year} {2004})\BibitemShut {NoStop}%
	\bibitem [{\citenamefont {Shankar}(2017)}]{Shankar_Book}%
	\BibitemOpen
	\bibfield  {author} {\bibinfo {author} {\bibfnamefont {R.}~\bibnamefont
			{Shankar}},\ }\href@noop {} {\emph {\bibinfo {title} {Quantum Field Theory
				and Condensed Matter: An Introduction}}}\ (\bibinfo  {publisher} {Cambridge
		University Press},\ \bibinfo {year} {2017})\BibitemShut {NoStop}%
	\bibitem [{\citenamefont {Qi}\ \emph {et~al.}(2008)\citenamefont {Qi},
		\citenamefont {Hughes},\ and\ \citenamefont {Zhang}}]{Qi2008}%
	\BibitemOpen
	\bibfield  {author} {\bibinfo {author} {\bibfnamefont {X.-L.}\ \bibnamefont
			{Qi}}, \bibinfo {author} {\bibfnamefont {T.~L.}\ \bibnamefont {Hughes}}, \
		and\ \bibinfo {author} {\bibfnamefont {S.-C.}\ \bibnamefont {Zhang}},\
	}\href@noop {} {\bibfield  {journal} {\bibinfo  {journal} {Nature Physics}\
		}\textbf {\bibinfo {volume} {4}},\ \bibinfo {pages} {273} (\bibinfo {year}
		{2008})}\BibitemShut {NoStop}%
	\bibitem [{\citenamefont {Peierls}(1936)}]{Peierls}%
	\BibitemOpen
	\bibfield  {author} {\bibinfo {author} {\bibfnamefont {R.}~\bibnamefont
			{Peierls}},\ }\href {\doibase 10.1017/S0305004100019174} {\bibfield
		{journal} {\bibinfo  {journal} {Mathematical Proceedings of the Cambridge
				Philosophical Society}\ }\textbf {\bibinfo {volume} {32}},\ \bibinfo {pages}
		{477} (\bibinfo {year} {1936})}\BibitemShut {NoStop}%
	\bibitem [{\citenamefont {Imry}\ and\ \citenamefont {Ma}(1975)}]{Imry1975}%
	\BibitemOpen
	\bibfield  {author} {\bibinfo {author} {\bibfnamefont {Y.}~\bibnamefont
			{Imry}}\ and\ \bibinfo {author} {\bibfnamefont {S.-k.}\ \bibnamefont {Ma}},\
	}\href {\doibase 10.1103/PhysRevLett.35.1399} {\bibfield  {journal} {\bibinfo
			{journal} {Phys. Rev. Lett.}\ }\textbf {\bibinfo {volume} {35}},\ \bibinfo
		{pages} {1399} (\bibinfo {year} {1975})}\BibitemShut {NoStop}%
	\bibitem [{\citenamefont {Pokrovsky}\ and\ \citenamefont
		{Talapov}(1979)}]{PokrovskyTalapov}%
	\BibitemOpen
	\bibfield  {author} {\bibinfo {author} {\bibfnamefont {V.~L.}\ \bibnamefont
			{Pokrovsky}}\ and\ \bibinfo {author} {\bibfnamefont {A.~L.}\ \bibnamefont
			{Talapov}},\ }\href {\doibase 10.1103/PhysRevLett.42.65} {\bibfield
		{journal} {\bibinfo  {journal} {Phys. Rev. Lett.}\ }\textbf {\bibinfo
			{volume} {42}},\ \bibinfo {pages} {65} (\bibinfo {year} {1979})}\BibitemShut
	{NoStop}%
	\bibitem [{\citenamefont {Giamarchi}\ and\ \citenamefont
		{Schulz}(1988)}]{Giamarchi1988}%
	\BibitemOpen
	\bibfield  {author} {\bibinfo {author} {\bibfnamefont {T.}~\bibnamefont
			{Giamarchi}}\ and\ \bibinfo {author} {\bibfnamefont {H.~J.}\ \bibnamefont
			{Schulz}},\ }\href {\doibase 10.1103/PhysRevB.37.325} {\bibfield  {journal}
		{\bibinfo  {journal} {Phys. Rev. B}\ }\textbf {\bibinfo {volume} {37}},\
		\bibinfo {pages} {325} (\bibinfo {year} {1988})}\BibitemShut {NoStop}%
	\bibitem [{\citenamefont {Kainaris}\ \emph {et~al.}(2014)\citenamefont
		{Kainaris}, \citenamefont {Gornyi}, \citenamefont {Carr},\ and\ \citenamefont
		{Mirlin}}]{Kainaris2014}%
	\BibitemOpen
	\bibfield  {author} {\bibinfo {author} {\bibfnamefont {N.}~\bibnamefont
			{Kainaris}}, \bibinfo {author} {\bibfnamefont {I.~V.}\ \bibnamefont
			{Gornyi}}, \bibinfo {author} {\bibfnamefont {S.~T.}\ \bibnamefont {Carr}}, \
		and\ \bibinfo {author} {\bibfnamefont {A.~D.}\ \bibnamefont {Mirlin}},\
	}\href {\doibase 10.1103/PhysRevB.90.075118} {\bibfield  {journal} {\bibinfo
			{journal} {Phys. Rev. B}\ }\textbf {\bibinfo {volume} {90}},\ \bibinfo
		{pages} {075118} (\bibinfo {year} {2014})}\BibitemShut {NoStop}%
	\bibitem [{\citenamefont {Fiete}\ \emph {et~al.}(2006)\citenamefont {Fiete},
		\citenamefont {Le~Hur},\ and\ \citenamefont {Balents}}]{Fiete2006}%
	\BibitemOpen
	\bibfield  {author} {\bibinfo {author} {\bibfnamefont {G.~A.}\ \bibnamefont
			{Fiete}}, \bibinfo {author} {\bibfnamefont {K.}~\bibnamefont {Le~Hur}}, \
		and\ \bibinfo {author} {\bibfnamefont {L.}~\bibnamefont {Balents}},\ }\href
	{\doibase 10.1103/PhysRevB.73.165104} {\bibfield  {journal} {\bibinfo
			{journal} {Phys. Rev. B}\ }\textbf {\bibinfo {volume} {73}},\ \bibinfo
		{pages} {165104} (\bibinfo {year} {2006})}\BibitemShut {NoStop}%
	\bibitem [{\citenamefont {Chou}\ \emph {et~al.}(2015)\citenamefont {Chou},
		\citenamefont {Levchenko},\ and\ \citenamefont {Foster}}]{Chou2015}%
	\BibitemOpen
	\bibfield  {author} {\bibinfo {author} {\bibfnamefont {Y.-Z.}\ \bibnamefont
			{Chou}}, \bibinfo {author} {\bibfnamefont {A.}~\bibnamefont {Levchenko}}, \
		and\ \bibinfo {author} {\bibfnamefont {M.~S.}\ \bibnamefont {Foster}},\
	}\href {\doibase 10.1103/PhysRevLett.115.186404} {\bibfield  {journal}
		{\bibinfo  {journal} {Phys. Rev. Lett.}\ }\textbf {\bibinfo {volume} {115}},\
		\bibinfo {pages} {186404} (\bibinfo {year} {2015})}\BibitemShut {NoStop}%
	\bibitem [{\citenamefont {Ziani}\ \emph {et~al.}(2015)\citenamefont {Ziani},
		\citenamefont {Cr\'epin},\ and\ \citenamefont {Trauzettel}}]{Ziani2015}%
	\BibitemOpen
	\bibfield  {author} {\bibinfo {author} {\bibfnamefont {N.~T.}\ \bibnamefont
			{Ziani}}, \bibinfo {author} {\bibfnamefont {F.}~\bibnamefont {Cr\'epin}}, \
		and\ \bibinfo {author} {\bibfnamefont {B.}~\bibnamefont {Trauzettel}},\
	}\href {\doibase 10.1103/PhysRevLett.115.206402} {\bibfield  {journal}
		{\bibinfo  {journal} {Phys. Rev. Lett.}\ }\textbf {\bibinfo {volume} {115}},\
		\bibinfo {pages} {206402} (\bibinfo {year} {2015})}\BibitemShut {NoStop}%
	\bibitem [{\citenamefont {Haldane}(1981)}]{Haldane1981}%
	\BibitemOpen
	\bibfield  {author} {\bibinfo {author} {\bibfnamefont {F.}~\bibnamefont
			{Haldane}},\ }\href@noop {} {\bibfield  {journal} {\bibinfo  {journal}
			{Journal of Physics C: Solid State Physics}\ }\textbf {\bibinfo {volume}
			{14}},\ \bibinfo {pages} {2585} (\bibinfo {year} {1981})}\BibitemShut
	{NoStop}%
	\bibitem [{\citenamefont {Bocquet}(1999)}]{Bocquet1999}%
	\BibitemOpen
	\bibfield  {author} {\bibinfo {author} {\bibfnamefont {M.}~\bibnamefont
			{Bocquet}},\ }\href@noop {} {\bibfield  {journal} {\bibinfo  {journal}
			{Nuclear Physics B}\ }\textbf {\bibinfo {volume} {546}},\ \bibinfo {pages}
		{621} (\bibinfo {year} {1999})}\BibitemShut {NoStop}%
	\bibitem [{\citenamefont {Fukuyama}\ and\ \citenamefont
		{Lee}(1978)}]{Fukuyama1978}%
	\BibitemOpen
	\bibfield  {author} {\bibinfo {author} {\bibfnamefont {H.}~\bibnamefont
			{Fukuyama}}\ and\ \bibinfo {author} {\bibfnamefont {P.~A.}\ \bibnamefont
			{Lee}},\ }\href {\doibase 10.1103/PhysRevB.17.535} {\bibfield  {journal}
		{\bibinfo  {journal} {Phys. Rev. B}\ }\textbf {\bibinfo {volume} {17}},\
		\bibinfo {pages} {535} (\bibinfo {year} {1978})}\BibitemShut {NoStop}%
	\bibitem [{\citenamefont {Narayan}\ and\ \citenamefont
		{Fisher}(1992)}]{Narayan1992}%
	\BibitemOpen
	\bibfield  {author} {\bibinfo {author} {\bibfnamefont {O.}~\bibnamefont
			{Narayan}}\ and\ \bibinfo {author} {\bibfnamefont {D.~S.}\ \bibnamefont
			{Fisher}},\ }\href {\doibase 10.1103/PhysRevB.46.11520} {\bibfield  {journal}
		{\bibinfo  {journal} {Phys. Rev. B}\ }\textbf {\bibinfo {volume} {46}},\
		\bibinfo {pages} {11520} (\bibinfo {year} {1992})}\BibitemShut {NoStop}%
	\bibitem [{\citenamefont {Giamarchi}\ and\ \citenamefont
		{Le~Doussal}(1996)}]{Giamarchi1996}%
	\BibitemOpen
	\bibfield  {author} {\bibinfo {author} {\bibfnamefont {T.}~\bibnamefont
			{Giamarchi}}\ and\ \bibinfo {author} {\bibfnamefont {P.}~\bibnamefont
			{Le~Doussal}},\ }\href {\doibase 10.1103/PhysRevLett.76.3408} {\bibfield
		{journal} {\bibinfo  {journal} {Phys. Rev. Lett.}\ }\textbf {\bibinfo
			{volume} {76}},\ \bibinfo {pages} {3408} (\bibinfo {year}
		{1996})}\BibitemShut {NoStop}%
	\bibitem [{\citenamefont {Balents}\ \emph {et~al.}(1998)\citenamefont
		{Balents}, \citenamefont {Marchetti},\ and\ \citenamefont
		{Radzihovsky}}]{Balents1998}%
	\BibitemOpen
	\bibfield  {author} {\bibinfo {author} {\bibfnamefont {L.}~\bibnamefont
			{Balents}}, \bibinfo {author} {\bibfnamefont {M.~C.}\ \bibnamefont
			{Marchetti}}, \ and\ \bibinfo {author} {\bibfnamefont {L.}~\bibnamefont
			{Radzihovsky}},\ }\href {\doibase 10.1103/PhysRevB.57.7705} {\bibfield
		{journal} {\bibinfo  {journal} {Phys. Rev. B}\ }\textbf {\bibinfo {volume}
			{57}},\ \bibinfo {pages} {7705} (\bibinfo {year} {1998})}\BibitemShut
	{NoStop}%
	\bibitem [{\citenamefont {Gor'kov}\ and\ \citenamefont
		{Gr{\"u}ner}(2012)}]{Gorkov_cdw_book}%
	\BibitemOpen
	\bibfield  {author} {\bibinfo {author} {\bibfnamefont {L.~P.}\ \bibnamefont
			{Gor'kov}}\ and\ \bibinfo {author} {\bibfnamefont {G.}~\bibnamefont
			{Gr{\"u}ner}},\ }\href@noop {} {\emph {\bibinfo {title} {Charge density waves
				in solids}}},\ Vol.~\bibinfo {volume} {25}\ (\bibinfo  {publisher}
	{Elsevier},\ \bibinfo {year} {2012})\BibitemShut {NoStop}%
	\bibitem [{\citenamefont {Imry}(2002)}]{Imry2002}%
	\BibitemOpen
	\bibfield  {author} {\bibinfo {author} {\bibfnamefont {Y.}~\bibnamefont
			{Imry}},\ }\href@noop {} {\emph {\bibinfo {title} {Introduction to mesoscopic
				physics}}}\ (\bibinfo  {publisher} {Oxford University Press on Demand},\
	\bibinfo {year} {2002})\BibitemShut {NoStop}%
	\bibitem [{\citenamefont {Khemani}\ \emph {et~al.}(2015)\citenamefont
		{Khemani}, \citenamefont {Nandkishore},\ and\ \citenamefont
		{Sondhi}}]{Khemani2015}%
	\BibitemOpen
	\bibfield  {author} {\bibinfo {author} {\bibfnamefont {V.}~\bibnamefont
			{Khemani}}, \bibinfo {author} {\bibfnamefont {R.}~\bibnamefont
			{Nandkishore}}, \ and\ \bibinfo {author} {\bibfnamefont {S.~L.}\ \bibnamefont
			{Sondhi}},\ }\href {\doibase 10.1038/nphys3344} {\bibfield  {journal}
		{\bibinfo  {journal} {Nat. Phys.}\ }\textbf {\bibinfo {volume} {11}},\
		\bibinfo {pages} {560} (\bibinfo {year} {2015})}\BibitemShut {NoStop}%
	\bibitem [{\citenamefont {Maciejko}\ \emph {et~al.}(2010)\citenamefont
		{Maciejko}, \citenamefont {Qi},\ and\ \citenamefont {Zhang}}]{Maciejko2010}%
	\BibitemOpen
	\bibfield  {author} {\bibinfo {author} {\bibfnamefont {J.}~\bibnamefont
			{Maciejko}}, \bibinfo {author} {\bibfnamefont {X.-L.}\ \bibnamefont {Qi}}, \
		and\ \bibinfo {author} {\bibfnamefont {S.-C.}\ \bibnamefont {Zhang}},\ }\href
	{\doibase 10.1103/PhysRevB.82.155310} {\bibfield  {journal} {\bibinfo
			{journal} {Phys. Rev. B}\ }\textbf {\bibinfo {volume} {82}},\ \bibinfo
		{pages} {155310} (\bibinfo {year} {2010})}\BibitemShut {NoStop}%
	\bibitem [{\citenamefont {Beenakker}(1997)}]{Beenakker1997}%
	\BibitemOpen
	\bibfield  {author} {\bibinfo {author} {\bibfnamefont {C.~W.~J.}\
			\bibnamefont {Beenakker}},\ }\href {\doibase 10.1103/RevModPhys.69.731}
	{\bibfield  {journal} {\bibinfo  {journal} {Rev. Mod. Phys.}\ }\textbf
		{\bibinfo {volume} {69}},\ \bibinfo {pages} {731} (\bibinfo {year}
		{1997})}\BibitemShut {NoStop}%
	\bibitem [{\citenamefont {Maciejko}\ \emph {et~al.}(2009)\citenamefont
		{Maciejko}, \citenamefont {Liu}, \citenamefont {Oreg}, \citenamefont {Qi},
		\citenamefont {Wu},\ and\ \citenamefont {Zhang}}]{Maciejko2009}%
	\BibitemOpen
	\bibfield  {author} {\bibinfo {author} {\bibfnamefont {J.}~\bibnamefont
			{Maciejko}}, \bibinfo {author} {\bibfnamefont {C.}~\bibnamefont {Liu}},
		\bibinfo {author} {\bibfnamefont {Y.}~\bibnamefont {Oreg}}, \bibinfo {author}
		{\bibfnamefont {X.-L.}\ \bibnamefont {Qi}}, \bibinfo {author} {\bibfnamefont
			{C.}~\bibnamefont {Wu}}, \ and\ \bibinfo {author} {\bibfnamefont {S.-C.}\
			\bibnamefont {Zhang}},\ }\href {\doibase 10.1103/PhysRevLett.102.256803}
	{\bibfield  {journal} {\bibinfo  {journal} {Phys. Rev. Lett.}\ }\textbf
		{\bibinfo {volume} {102}},\ \bibinfo {pages} {256803} (\bibinfo {year}
		{2009})}\BibitemShut {NoStop}%
	\bibitem [{\citenamefont {V\"ayrynen}\ \emph {et~al.}(2016)\citenamefont
		{V\"ayrynen}, \citenamefont {Geissler},\ and\ \citenamefont
		{Glazman}}]{Vayrynen2016}%
	\BibitemOpen
	\bibfield  {author} {\bibinfo {author} {\bibfnamefont {J.~I.}\ \bibnamefont
			{V\"ayrynen}}, \bibinfo {author} {\bibfnamefont {F.}~\bibnamefont
			{Geissler}}, \ and\ \bibinfo {author} {\bibfnamefont {L.~I.}\ \bibnamefont
			{Glazman}},\ }\href {\doibase 10.1103/PhysRevB.93.241301} {\bibfield
		{journal} {\bibinfo  {journal} {Phys. Rev. B}\ }\textbf {\bibinfo {volume}
			{93}},\ \bibinfo {pages} {241301} (\bibinfo {year} {2016})}\BibitemShut
	{NoStop}%
	\bibitem [{\citenamefont {Teo}\ and\ \citenamefont {Kane}(2009)}]{Teo2009}%
	\BibitemOpen
	\bibfield  {author} {\bibinfo {author} {\bibfnamefont {J.~C.~Y.}\
			\bibnamefont {Teo}}\ and\ \bibinfo {author} {\bibfnamefont {C.~L.}\
			\bibnamefont {Kane}},\ }\href {\doibase 10.1103/PhysRevB.79.235321}
	{\bibfield  {journal} {\bibinfo  {journal} {Phys. Rev. B}\ }\textbf {\bibinfo
			{volume} {79}},\ \bibinfo {pages} {235321} (\bibinfo {year}
		{2009})}\BibitemShut {NoStop}%
	\bibitem [{\citenamefont {Li}\ \emph {et~al.}(2018)\citenamefont {Li},
		\citenamefont {Zhang},\ and\ \citenamefont {Shen}}]{Li2017_Hidden_edge}%
	\BibitemOpen
	\bibfield  {author} {\bibinfo {author} {\bibfnamefont {C.-A.}\ \bibnamefont
			{Li}}, \bibinfo {author} {\bibfnamefont {S.-B.}\ \bibnamefont {Zhang}}, \
		and\ \bibinfo {author} {\bibfnamefont {S.-Q.}\ \bibnamefont {Shen}},\ }\href
	{\doibase 10.1103/PhysRevB.97.045420} {\bibfield  {journal} {\bibinfo
			{journal} {Phys. Rev. B}\ }\textbf {\bibinfo {volume} {97}},\ \bibinfo
		{pages} {045420} (\bibinfo {year} {2018})}\BibitemShut {NoStop}%
	\bibitem [{\citenamefont {Skolasinski}\ \emph {et~al.}(2017)\citenamefont
		{Skolasinski}, \citenamefont {Pikulin}, \citenamefont {Alicea},\ and\
		\citenamefont {Wimmer}}]{Skolasinski2017}%
	\BibitemOpen
	\bibfield  {author} {\bibinfo {author} {\bibfnamefont {R.}~\bibnamefont
			{Skolasinski}}, \bibinfo {author} {\bibfnamefont {D.~I.}\ \bibnamefont
			{Pikulin}}, \bibinfo {author} {\bibfnamefont {J.}~\bibnamefont {Alicea}}, \
		and\ \bibinfo {author} {\bibfnamefont {M.}~\bibnamefont {Wimmer}},\
	}\href@noop {} {\bibfield  {journal} {\bibinfo  {journal} {arXiv preprint
				arXiv:1709.04830}\ } (\bibinfo {year} {2017})}\BibitemShut {NoStop}%
	\bibitem [{\citenamefont {Pikulin}\ \emph {et~al.}(2014)\citenamefont
		{Pikulin}, \citenamefont {Hyart}, \citenamefont {Mi}, \citenamefont
		{Tworzyd\l{}o}, \citenamefont {Wimmer},\ and\ \citenamefont
		{Beenakker}}]{Pikulin2014}%
	\BibitemOpen
	\bibfield  {author} {\bibinfo {author} {\bibfnamefont {D.~I.}\ \bibnamefont
			{Pikulin}}, \bibinfo {author} {\bibfnamefont {T.}~\bibnamefont {Hyart}},
		\bibinfo {author} {\bibfnamefont {S.}~\bibnamefont {Mi}}, \bibinfo {author}
		{\bibfnamefont {J.}~\bibnamefont {Tworzyd\l{}o}}, \bibinfo {author}
		{\bibfnamefont {M.}~\bibnamefont {Wimmer}}, \ and\ \bibinfo {author}
		{\bibfnamefont {C.~W.~J.}\ \bibnamefont {Beenakker}},\ }\href {\doibase
		10.1103/PhysRevB.89.161403} {\bibfield  {journal} {\bibinfo  {journal} {Phys.
				Rev. B}\ }\textbf {\bibinfo {volume} {89}},\ \bibinfo {pages} {161403}
		(\bibinfo {year} {2014})}\BibitemShut {NoStop}%
	\bibitem [{\citenamefont {Hu}\ \emph {et~al.}(2016)\citenamefont {Hu},
		\citenamefont {Xu}, \citenamefont {Zhang},\ and\ \citenamefont
		{Zhou}}]{Hu2016}%
	\BibitemOpen
	\bibfield  {author} {\bibinfo {author} {\bibfnamefont {L.-H.}\ \bibnamefont
			{Hu}}, \bibinfo {author} {\bibfnamefont {D.-H.}\ \bibnamefont {Xu}}, \bibinfo
		{author} {\bibfnamefont {F.-C.}\ \bibnamefont {Zhang}}, \ and\ \bibinfo
		{author} {\bibfnamefont {Y.}~\bibnamefont {Zhou}},\ }\href {\doibase
		10.1103/PhysRevB.94.085306} {\bibfield  {journal} {\bibinfo  {journal} {Phys.
				Rev. B}\ }\textbf {\bibinfo {volume} {94}},\ \bibinfo {pages} {085306}
		(\bibinfo {year} {2016})}\BibitemShut {NoStop}%
	\bibitem [{\citenamefont {V\"ayrynen}\ \emph {et~al.}(2013)\citenamefont
		{V\"ayrynen}, \citenamefont {Goldstein},\ and\ \citenamefont
		{Glazman}}]{Vayrnen2013}%
	\BibitemOpen
	\bibfield  {author} {\bibinfo {author} {\bibfnamefont {J.~I.}\ \bibnamefont
			{V\"ayrynen}}, \bibinfo {author} {\bibfnamefont {M.}~\bibnamefont
			{Goldstein}}, \ and\ \bibinfo {author} {\bibfnamefont {L.~I.}\ \bibnamefont
			{Glazman}},\ }\href {\doibase 10.1103/PhysRevLett.110.216402} {\bibfield
		{journal} {\bibinfo  {journal} {Phys. Rev. Lett.}\ }\textbf {\bibinfo
			{volume} {110}},\ \bibinfo {pages} {216402} (\bibinfo {year}
		{2013})}\BibitemShut {NoStop}%
	\bibitem [{\citenamefont {V\"ayrynen}\ \emph {et~al.}(2014)\citenamefont
		{V\"ayrynen}, \citenamefont {Goldstein}, \citenamefont {Gefen},\ and\
		\citenamefont {Glazman}}]{Vayrnen2014}%
	\BibitemOpen
	\bibfield  {author} {\bibinfo {author} {\bibfnamefont {J.~I.}\ \bibnamefont
			{V\"ayrynen}}, \bibinfo {author} {\bibfnamefont {M.}~\bibnamefont
			{Goldstein}}, \bibinfo {author} {\bibfnamefont {Y.}~\bibnamefont {Gefen}}, \
		and\ \bibinfo {author} {\bibfnamefont {L.~I.}\ \bibnamefont {Glazman}},\
	}\href {\doibase 10.1103/PhysRevB.90.115309} {\bibfield  {journal} {\bibinfo
			{journal} {Phys. Rev. B}\ }\textbf {\bibinfo {volume} {90}},\ \bibinfo
		{pages} {115309} (\bibinfo {year} {2014})}\BibitemShut {NoStop}%
	\bibitem [{\citenamefont {Tanaka}\ \emph {et~al.}(2011)\citenamefont {Tanaka},
		\citenamefont {Furusaki},\ and\ \citenamefont {Matveev}}]{Tanaka2011}%
	\BibitemOpen
	\bibfield  {author} {\bibinfo {author} {\bibfnamefont {Y.}~\bibnamefont
			{Tanaka}}, \bibinfo {author} {\bibfnamefont {A.}~\bibnamefont {Furusaki}}, \
		and\ \bibinfo {author} {\bibfnamefont {K.~A.}\ \bibnamefont {Matveev}},\
	}\href {\doibase 10.1103/PhysRevLett.106.236402} {\bibfield  {journal}
		{\bibinfo  {journal} {Phys. Rev. Lett.}\ }\textbf {\bibinfo {volume} {106}},\
		\bibinfo {pages} {236402} (\bibinfo {year} {2011})}\BibitemShut {NoStop}%
	\bibitem [{\citenamefont {Altshuler}\ \emph {et~al.}(2013)\citenamefont
		{Altshuler}, \citenamefont {Aleiner},\ and\ \citenamefont
		{Yudson}}]{Altshuler2013}%
	\BibitemOpen
	\bibfield  {author} {\bibinfo {author} {\bibfnamefont {B.~L.}\ \bibnamefont
			{Altshuler}}, \bibinfo {author} {\bibfnamefont {I.~L.}\ \bibnamefont
			{Aleiner}}, \ and\ \bibinfo {author} {\bibfnamefont {V.~I.}\ \bibnamefont
			{Yudson}},\ }\href {\doibase 10.1103/PhysRevLett.111.086401} {\bibfield
		{journal} {\bibinfo  {journal} {Phys. Rev. Lett.}\ }\textbf {\bibinfo
			{volume} {111}},\ \bibinfo {pages} {086401} (\bibinfo {year}
		{2013})}\BibitemShut {NoStop}%
	\bibitem [{\citenamefont {Hsu}\ \emph {et~al.}(2017)\citenamefont {Hsu},
		\citenamefont {Stano}, \citenamefont {Klinovaja},\ and\ \citenamefont
		{Loss}}]{Hsu2017}%
	\BibitemOpen
	\bibfield  {author} {\bibinfo {author} {\bibfnamefont {C.-H.}\ \bibnamefont
			{Hsu}}, \bibinfo {author} {\bibfnamefont {P.}~\bibnamefont {Stano}}, \bibinfo
		{author} {\bibfnamefont {J.}~\bibnamefont {Klinovaja}}, \ and\ \bibinfo
		{author} {\bibfnamefont {D.}~\bibnamefont {Loss}},\ }\href {\doibase
		10.1103/PhysRevB.96.081405} {\bibfield  {journal} {\bibinfo  {journal} {Phys.
				Rev. B}\ }\textbf {\bibinfo {volume} {96}},\ \bibinfo {pages} {081405}
		(\bibinfo {year} {2017})}\BibitemShut {NoStop}%
	\bibitem [{\citenamefont {Foster}\ \emph {et~al.}(2010)\citenamefont {Foster},
		\citenamefont {Yuzbashyan},\ and\ \citenamefont {Altshuler}}]{Foster2010}%
	\BibitemOpen
	\bibfield  {author} {\bibinfo {author} {\bibfnamefont {M.~S.}\ \bibnamefont
			{Foster}}, \bibinfo {author} {\bibfnamefont {E.~A.}\ \bibnamefont
			{Yuzbashyan}}, \ and\ \bibinfo {author} {\bibfnamefont {B.~L.}\ \bibnamefont
			{Altshuler}},\ }\href {\doibase 10.1103/PhysRevLett.105.135701} {\bibfield
		{journal} {\bibinfo  {journal} {Phys. Rev. Lett.}\ }\textbf {\bibinfo
			{volume} {105}},\ \bibinfo {pages} {135701} (\bibinfo {year}
		{2010})}\BibitemShut {NoStop}%
	\bibitem [{\citenamefont {Matveev}\ \emph {et~al.}(1993)\citenamefont
		{Matveev}, \citenamefont {Yue},\ and\ \citenamefont {Glazman}}]{Matveev1993}%
	\BibitemOpen
	\bibfield  {author} {\bibinfo {author} {\bibfnamefont {K.~A.}\ \bibnamefont
			{Matveev}}, \bibinfo {author} {\bibfnamefont {D.}~\bibnamefont {Yue}}, \ and\
		\bibinfo {author} {\bibfnamefont {L.~I.}\ \bibnamefont {Glazman}},\ }\href
	{\doibase 10.1103/PhysRevLett.71.3351} {\bibfield  {journal} {\bibinfo
			{journal} {Phys. Rev. Lett.}\ }\textbf {\bibinfo {volume} {71}},\ \bibinfo
		{pages} {3351} (\bibinfo {year} {1993})}\BibitemShut {NoStop}%
	\bibitem [{\citenamefont {Garst}\ \emph {et~al.}(2008)\citenamefont {Garst},
		\citenamefont {Novikov}, \citenamefont {Stern},\ and\ \citenamefont
		{Glazman}}]{Garst2008}%
	\BibitemOpen
	\bibfield  {author} {\bibinfo {author} {\bibfnamefont {M.}~\bibnamefont
			{Garst}}, \bibinfo {author} {\bibfnamefont {D.~S.}\ \bibnamefont {Novikov}},
		\bibinfo {author} {\bibfnamefont {A.}~\bibnamefont {Stern}}, \ and\ \bibinfo
		{author} {\bibfnamefont {L.~I.}\ \bibnamefont {Glazman}},\ }\href {\doibase
		10.1103/PhysRevB.77.035128} {\bibfield  {journal} {\bibinfo  {journal} {Phys.
				Rev. B}\ }\textbf {\bibinfo {volume} {77}},\ \bibinfo {pages} {035128}
		(\bibinfo {year} {2008})}\BibitemShut {NoStop}%
\end{thebibliography}

%%%%%%%%%%%%%%


%%%%%%%%%%%%%%
	
\end{document}